%% file: TTO_hardening.tex
\documentclass[leqno,12pt,a4paper]{article}

\usepackage{setspace}
\usepackage{latexsym}
\usepackage{amsmath}
\usepackage{amssymb}
\usepackage{mathtools}
\usepackage{bbm}
\usepackage{amsthm}
\usepackage[misc]{ifsym}

\usepackage[]{algorithm,algpseudocode}
\algnewcommand\algorithmicforeach{\textbf{for each}}
\algdef{S}[FOR]{ForEach}[1]{\algorithmicforeach\ #1\ \algorithmicdo}

\usepackage{graphicx}
\usepackage{epsfig}
\usepackage{subcaption}
\usepackage{tabularx}
\usepackage{array}
\usepackage{multirow} 
\newcolumntype{L}[1]{>{\raggedright\arraybackslash}p{#1}} 
\newcolumntype{C}[1]{>{\centering\arraybackslash}p{#1}} 
\newcolumntype{R}[1]{>{\raggedleft\arraybackslash}p{#1}} 
\newcolumntype{M}[1]{>{\centering\arraybackslash}m{#1}} 

\allowdisplaybreaks
\usepackage{tikz}
\usetikzlibrary{shapes,arrows,patterns,backgrounds,fit,arrows.meta,shapes.misc}
\tikzset{
  font=\fontfamily{phv}\selectfont\small,
}
\usepackage{pgfplots}
\pgfplotsset{
  compat=newest,
  width=0.65\textwidth,
  height=0.4\textwidth,
  tick align=outside,
  tick pos=left,
  scaled ticks=false,
  grid=major,
  grid style={dotted},
  every tick label/.append style={font=\fontfamily{phv}\selectfont\scriptsize},
}
\pgfkeys{/pgf/number format/.cd,
  1000 sep={\,}, 
  min exponent for 1000 sep=4,
  assume math mode 
}
\tikzset{cross/.style={cross out, draw, 
         minimum size=2*(#1-\pgflinewidth), 
         inner sep=0pt, outer sep=0pt}}

\usepackage{siunitx}
\usepackage{makecell}

\usepackage{placeins} 

\usepackage[
    pdfpagemode = UseNone,  
    pdfstartview={FitH},
    bookmarksnumbered, 
    bookmarksopenlevel=section,
    hidelinks, 
    linktoc=all, 
    pdftitle={Thermodynamic topology optimization including plasticity},
    pdfauthor={Miriam Kick, Philipp Junker},
    pdftex, 
]{hyperref}

\newcommand{\boldface}[1]{\boldsymbol{#1}}
\newcommand{\bfa}{\boldface{a}}
\newcommand{\bfb}{\boldface{b}}
\newcommand{\bfc}{\boldface{c}}

\newcommand{\bff}{\boldface{f}}

\newcommand{\bfn}{\boldface{n}}

\newcommand{\bfp}{\boldface{p}}

\newcommand{\bfr}{\boldface{r}}
\newcommand{\bfs}{\boldface{s}}
\newcommand{\bft}{\boldface{t}}
\newcommand{\bfu}{\boldface{u}}

\newcommand{\bfx}{\boldface{x}}
\newcommand{\bfy}{\boldface{y}}

\newcommand{\bfA}{\boldface{A}}
\newcommand{\bfB}{\boldface{B}}
\newcommand{\bfC}{\boldface{C}}

\newcommand{\bfI}{\boldface{I}}

\newcommand{\bfK}{\boldface{K}}

\newcommand{\bfN}{\boldface{N}}

\newcommand{\bfP}{\boldface{P}}

\newcommand{\bfT}{\boldface{T}}

\newcommand{\bfalpha}{\boldsymbol{\alpha}}

\newcommand{\bfsigma}{\boldsymbol{\sigma}}

\newcommand{\bfeps}{\boldsymbol{\varepsilon}}
\newcommand{\calC}{\mathcal{C}}
\newcommand{\calD}{\mathcal{D}}

\newcommand{\calF}{\mathcal{F}}
\newcommand{\calG}{\mathcal{G}}
\newcommand{\calH}{\mathcal{H}}

\newcommand{\calR}{\mathcal{R}}

\usepackage{bbold}
\newcommand{\dsA}{\mathbb{A}}

\newcommand{\dsC}{\mathbb{C}}
\newcommand{\dsD}{\mathbb{D}}
\newcommand{\dsE}{\mathbb{E}}

\newcommand{\dsI}{\mathbb{I}}

\newcommand{\dsP}{\mathbb{P}}

\newcommand{\be}{\begin{equation}}
\newcommand{\ee}{\end{equation}}
\newcommand{\bea}{\begin{eqnarray}}
\newcommand{\eea}{\end{eqnarray}}
\newcommand{\bes}{\begin{equation*}}
\newcommand{\ees}{\end{equation*}}
\newcommand{\beas}{\begin{eqnarray*}}
\newcommand{\eeas}{\end{eqnarray*}}
\newcommand{\D}{\displaystyle}

\newcommand{\tr}[1]{\mathrm{tr} \, #1}

\newcommand{\pf}[2]{\frac{\partial #1}{\partial #2}}
\newcommand{\comp}[1]{{(#1)}}

\newcommand{\diss}{\mathrm{diss}}
\newcommand{\bfpdiss}{\bfp^\diss}

\newcommand{\lap}{\mathop{}\!\mathbin\bigtriangleup}
\newcommand{\dd}{\ \mathrm{d}}

\newcommand{\norm}[1]{\left\lVert#1\right\rVert}

\newcommand{\bfepsp}{\boldsymbol{\varepsilon}^{\mathrm{p}}}

\newcommand{\epsYexp}{\varepsilon_\mathrm{exp}^\mathrm{Y}}

\newcommand{\ppf}[2]{\frac{\partial^2 #1}{\partial #2 \partial #2}}
\newcommand{\tf}[2]{\frac{\mathrm{d} #1}{\mathrm{d} #2}}

\newcommand{\bfsigmadev}{\bfsigma^{\mathrm{dev}}}

\newcommand{\sigmaY}{\sigma^\mathrm{Y}}
\newcommand{\sigmaYexp}{\sigmaY_\mathrm{exp}}

\newcommand{\relStructVol}{v_{0}}
\newcommand{\rH}{r \left(\norm{\bfepsp} \right)}
\newcommand{\dr}{\pf{r}{\bfepsp}}
\newcommand{\ddr}{\frac{\partial^2 r}{\partial \bfepsp \partial \bfepsp}}
\newcommand{\dsEChi}{\chi^3 \, \dsE_0}
\newcommand{\drivingForce}{p}
\newcommand{\dsdepsp}{\bfP}
\newcommand{\dsdeps}{\bfT}

\immediate\pdfobj stream attr{/N 3} file{sRGB.icc}
\pdfcatalog{%
  /OutputIntents [
    <<
      /Type /OutputIntent
      /S /GTS_PDFA1
      /DestOutputProfile \the\pdflastobj\space 0 R
      /OutputConditionIdentifier (sRGB)
      /Info (sRGB)
    >>
    <<
      /Type /OutputIntent
      /S /GTS_PDFX
      /DestOutputProfile \the\pdflastobj\space 0 R
      /OutputConditionIdentifier (sRGB)
      /Info (sRGB)
    >>
  ]
}

\definecolor{LUHblue}{RGB}{0,80,155}
\definecolor{LUHblue40}{RGB}{153,185,216}
\definecolor{LUHblue20}{RGB}{204,220,235}
\definecolor{LUHblack40}{RGB}{153,153,153}
\definecolor{LUHblack20}{RGB}{204,204,204}
\definecolor{LUHgreen}{RGB}{200,211,23}
\definecolor{IKMgreen}{RGB}{175,218,0}
\definecolor{lightgray}{gray}{0.92}
\definecolor{greenSPI}{RGB}{130,172,30}
\definecolor{greenSPH1}{RGB}{85,197,104}
\definecolor{greenSPH2}{RGB}{0,140,114}
\definecolor{grayE}{RGB}{82,82,81}
\definecolor{blueCP}{RGB}{0,80,155}

\definecolor{applegreen}{rgb}{0.55, 0.71, 0.0}

\makeatletter
\renewenvironment{cases}[1][l]{\matrix@check\cases\env@cases{#1}}{\endarray\right.}
\def\env@cases#1{%
  \let\@ifnextchar\new@ifnextchar
  \left\lbrace\def\arraystretch{1.2}%
  \array{@{}#1@{\quad}l@{}}}
\makeatother

\usepackage{booktabs}

\RequirePackage{relsize}
\newfont{\Sf}{cmssbx10 scaled 2074}

\newcommand{\AssemargPJ}[2]{{
  \underset{#1}{\overset{\mathsmaller{#2}}%
  {\raisebox{-0.5ex}{\mbox{\Sf A}}}}\;%
  }}

\newcommand{\figref}[1]{Fig.~\ref{#1}}
\newcommand{\tabref}[1]{Tab.~\ref{#1}}
\newcommand{\secref}[1]{Sec.~\ref{#1}}
\newcommand{\appref}[1]{App.~\ref{#1}}
\newcommand{\algoref}[1]{Alg.~\ref{#1}}

\newcommand{\eg}{e.\,g.}
\newcommand{\ie}{i.\,e.}
\newcommand{\etal}{\textit{et al.}}

\begin{document}
\begin{singlespace}
\title{Thermodynamic topology optimization for hardening materials}
\date{} 
\maketitle
{\large
\noindent{Miriam Kick}, Philipp Junker\\[0.5mm]
}
Leibniz University Hannover, Institute of Continuum Mechanics, Hannover, Germany\\[2mm]
{
Corresponding author:\\[0.5mm]
Philipp Junker, \color{LUHblue}{\Letter \hskip 2mm junker@ikm.uni-hannover.de}
}
\end{singlespace}

\addcontentsline{toc}{section}{Abstract}
\section*{Abstract}
Topology optimization is an important basis for the design of components.
Here, the optimal structure is found within a design space subject to boundary conditions.
Thereby, the specific material law has a strong impact on the final design.
An important kind of material behavior is hardening: then a, for instance, linear-elastic structure is not optimal if plastic deformation will be induced by the loads.
Since hardening behavior has a remarkable impact on the resultant stress field, it needs to be accounted for during topology optimization.
In this contribution, we present an extension of the thermodynamic topology optimization that accounts for this non-linear material behavior due to the evolution of plastic strains.
For this purpose, we develop a novel surrogate model that allows to compute the plastic strain tensor corresponding to the current structure design for arbitrary hardening behavior.
We show the agreement of the model with the classic plasticity model for monotonic loading.
Furthermore, we demonstrate the interaction of the topology optimization for hardening material behavior results in structural changes.\\[2mm]

\noindent
\textbf{Keywords:}\\ Thermodynamic topology optimization; surrogate model for hardening materials; evolutionary approach; arbitrary hardening.

\section{Introduction}
Engineers are always looking for structures that meet the specific requirements in an optimal way.
One possibility for finding these structures is provided by optimization schemes which are classified as follows: i) improving the principal idea, ii) modifying the material, iii) thickness dimensioning, iv) optimization of shape and v) optimization of topology \cite{schumacher2013Optimierung,harzheim2008strukturoptimierung}.
Herein, the optimization scheme that demands the minimum amount of restrictions is given by topology optimization.
The consideration of the real materials properties offers additional potential for the optimal design of components.
Therefore, it is important to account for the physical material behavior even during the process of topology optimization.

There are various variants of topology optimization available as, \eg{}, the optimization for temperature evolution, noise reduction, dynamic response, or structural stiffness.
All of these approaches have in common that the related physical balance laws, in most cases the balance of linear momentum, are solved along with a mathematical optimization problem which is given in terms of an objective function.
The most common objective is the minimization of compliance, \ie{}, the maximization of stiffness according to a target structure volume \cite{sigmund2013topology}.
Therefore, topology optimization determines the position and arrangement of material within a given design space and boundary conditions such that the objective function is minimized.
The topology of a structure can be parameterized via different approaches during the specific numerical investigation.
For the numerical solution of the physical balance law, \eg{}, the balance of linear momentum, usually the finite element method (FEM) is employed.
Consequently, the finite elements introduce a discretization of the design space, and it is thus most common to assign a density value for each discrete subvolume, \ie{}, for each finite element.
For this assignment, a variety of different numerical schemes has been developed among which the probably most popular is given by ``Solid Material with Penalization'' (SIMP) proposed by Bendsøe and Sigmund in \cite{Bendsoe1989Optimal,bendsoe2003Topology}.
The fundamental idea of SIMP is the introduction of a non-linear interpolation function between void and full material such that a black and white design is aspired due to the inherent non-convex total potential.
Further popular developments are overviewed in \cite{sigmund2013topology,Deaton2014ASurvey}.

In a series of papers, we aimed at contributing to the problem of topology optimization: by using thermodynamic extremal principles, evolutionary access to the problem of topology optimization has been presented for which we referred our method to as thermodynamic topology optimization (TTO).
It routes back to \cite{junker2015variational} while further important developments have been presented for the numerical treatment in \cite{jantos2019accurate} and for hyperelastic material in \cite{Junker2021hyperelastic}.
This topology optimization makes use of an extended Hamilton principle which is well-known in the context of material modeling, cf. \cite{junker2021extended}.
Thus, the great advantage of this method is that it is able to factor in complex material behavior into topology optimization.
The extended Hamilton functional is formulated and its stationarity conditions serve as update procedure for the evolution of the topology.
In this manner, no classic optimization problem is formulated.
Since the free energy function is part of the extended Hamilton functional, the result is very similar to classical schemes for topology optimization with the objective of minimization of compliance.
All constraints need to be formulated as part of the Hamilton functional and affect the governing system of equations.
Therefore, classic optimization algorithms are more flexible with independent definition of different objectives and constraints.
On the other hand, no additional optimization algorithm is needed for thermodynamic topology optimization.
In contrast to classic optimization, the complete optimization process results as smooth computation of a system of differential equations.
To this end, the relative density of this density-based approach is described by a transient partial differential equation (PDE) in which the local free energy density serves as source term.
Consequently, the material optimization problem is converted to an evolutionary problem.
The field equation for the topology results from the stationary condition of the extended Hamilton functional.
Additionally, evaluation of the functional results in the field equations for displacement and internal (state) variable which accounts for the (local) microstructure of the material.
From this follows that the extended Hamilton functional according to topology optimization also features to take any physically dissipative material behavior into account.

In context of accounting for a hardening material behavior during the optimization, direct access of complex non-linear material behavior within topology optimization is required.
In general, these approaches take hardening into account.
Plastic material behavior requires a thermodynamically and mathematically rigorous treatment due to its complexity.
The integration might be given by using classic plasticity models with the characteristic stress/strain diagram resulting in a hysteresis curve in cyclic loading.
Several successful examples are provided in the literature: a first approach to account for a classic elasto-plastic material model within an adaptive material topology optimization was proposed by Maute \etal{} \cite{Maute1998Adaptive}.
Approaches to determine plasticity by homogenization strategies are also possible, cf. \cite{Yuge1995Optimization}.
This is particularly interesting for plastic parts of composites \cite{Swan1998VoigtReuss}.
Further studies consider composites due to classic elasto-plastic formulations, \eg{}, Kato \etal{} by multiphase material optimization.
Topology optimization can also be used to compute optimized distributions of two materials within a composite structure with respect to different strength in tension and compression, \eg{}, steel and concrete, as stated by Bogomolny and Amir \cite{Bogomolny2012Conceptual}.
An even more complex material behavior can be considered by including anisotropic plastic materials into topology optimization, cf. \cite{Zhang2017Topology}.
Geometric non-linearities in plastic material behavior are also of interest.
Therefore, topology optimization based on finite strains can be considered for plasticity, for instance by Wallin \etal{} \cite{Wallin2016Topology}.
Russ and Waisman \cite{Russ2021ANovel} proposed an enhanced approach to optimize structures in context of high loading while buckling is not desired by separation of linear elastic buckling analysis and the small strain elastoplastic material.
Different load types, \eg{}, optimizing dynamically loaded structures while accounting for plastic material behavior \cite{Nakshatrala2015Topology}, are considered.
In case of cyclic loading due to elasto-plastic optimization, the importance of path-dependence is particularly evident and cannot be neglected, cf. \cite{Li2017Design}.
A different option was proposed by the consideration of damage, cf. \cite{Li2017Topology,Alberdi2017Topology}.
Usually, classic elasto-plastic material is modeled with respect to hardening, \eg{}, bilinearly with elasticity and linear hardening.
However, hardening can also be described for a pure loading case with a multilinear elasto-plastic model established by Yoon and Kim \cite{Yoon2007Topology}.
In other studies, topology and shape optimization with respect to elasto-plastic material behavior are combined, for instance by Schwarz \etal{} \cite{Schwarz2001Topology}.
For all such non-linear stress/strain relations, the topology optimization routine usually demands an additional optimization algorithm for convergence.
Here, one prominent possibility is provided by the ``method of moving asymptotes'' (MMA).

Unfortunately, the strategy of considering physical material models usually renders such optimization rather time-consuming: due to the local path-dependence, the physical loading process needs to be discretized with several time steps each of which demands the solution of the physical finite element problem.
The nested finite element simulations for the physical process and the topology optimization problem demand a remarkably higher amount of computation time.
To compensate this drawback, several strategies can be found which aim to directly include plasticity or selected elasto-plastic characteristics into the optimization process.
One possibility is to make use of an elastic model with local stress constraint as mentioned, \eg{}, by \cite{Duysinx1998Topology,Duysinx1999Topology,Bruggi2012Topology,Luo2012Topology}.
The definition of stress constraints ensures mechanical strength for high loading and they are similar to the yield stress according to ideal plasticity.
Another idea by Amir \cite{Amir2016StressConstrained} is to define a single global stress constraint within the formulation of the optimization problem to bypass the local calculation for each material point.
Another way is to account for the non-linear material behavior on a second, microscopic scale by developing a new model reduction technique which is proposed by Fritzen \etal{} \cite{Fritzen2016Topology}.
The optimization on the macroscopic scale is extended by Xia \etal{} \cite{Xia2017Evolutionary} for greater robustness by means of damping.
A special characteristic of this approach is the use of an evolutionary optimization method on the macroscopic scale.
The deformation theory of plasticity by Hencky \cite{Hencky1924ZurTheorie,Duester2001ThePVersion} shows the possibility of modelling a perfect plastic material behavior without path-dependence.
The stress/strain relation is modeled by a non-linear algebraic equation under usage of real material parameters.
Updating the material state has very low computational cost and only one finite element simulation is needed due to the path-independence.
Hencky plasticity is modeled on assumptions that make this model only reasonable under certain conditions, \eg{}, monotonic loading up to a damage threshold.
This requirement is fulfilled in optimization such that Hencky plasticity can be used, for instance, for shape optimization, cf. \cite{Maury2018ElastPlastic}.
There exist approaches for extending Hencky plasticity for hardening material behavior \cite{Hencky1933TheNewTheory}.
Unfortunately, these early ideas lack consistent physical reasoning, \eg{}, the introduced parameters are not measurable.
Furthermore, models have been developed which avoid the need of solving a physical tensor-valued evolution equation for the microstructural material behavior.
One approach is to make use of power law models, \eg{}, by Ramberg and Osgood \cite{RambergOsgood1943Description} who fit a 1D stress/strain curve by three parameters which are not directly related to real material parameters.
Unfortunately, the calculation of the Ramberg-Osgood parameters is a rather intricate task.  
The update of the material state on the 1D von Mises stress/strain relationship level results in a scalar-valued updated von Mises stress.
Accordingly, this scalar-valued von Mises stress needs to be further processed to recover a 3D tensor-valued stresses and the corresponding tensor-valued plastic strains.
The Ramberg-Osgood material law is, \eg{}, used for topology optimization with large deformations in \cite{Huang2008Topology}.
In addition, \eg{}, Zhao~et~al.~\cite{Zhao2019Material,Zhao2020Topology} developed a surrogate model as a fictitious non-linear elastic material model which asymptotically approximates a perfect elasto-plastic behavior.
They accounted for the von Mises criterion in \cite{Zhao2019Material} and also developed an approach valid for the Drucker-Prager criterion in \cite{Zhao2020Topology}.
Due to the absent path-dependence, the computation of the sensitivity is straight forward and only one finite element analysis needs to be computed for each iteration step.
Therefore, this approach has a remarkable disadvantage that the resulting stress/strain curve matches the similar classic elasto-plastic curve, even at a material point level only at the limit points.
Furthermore, there is no possibility to compute the plastic strain tensor which serves as thermodynamic state variable.

In this contribution, we aim at expanding the thermodynamic topology optimization such that it can be applied to hardening materials with a novel 3D~surrogate material model.
We propose the surrogate model to represent hardening behavior of materials that are loaded beyond their initial yield criterion.
Therefore, it should be possible to achieve any type of hardening behavior with respect to real material parameters.
Since hardening is a special case of plasticity, the surrogate model is based on a classic plastic material model which depend on a 3D plastic strain tensor.
The classic model is reduced by modifications resulting from the nature of optimization: we determine topology optimization results for the (maximal) external loading which is employed by \emph{monotonic} loading, \ie{}, no unloading or loading cycles can be taken into account.
To this end, we assume that plastic strains evolve "virtually" through changes in structure during evolutionary topology optimization and physical unloading does not occur.
Hence, a single loading point needs to be addressed in the novel stress/strain relation without dissipation and path-dependence.
Detailed explanations follow in \secref{sec:specificationDerivation} together with the definition of energy and constraint formulations.
Classic models account for all properties of complex plasticity behavior and are thus physically accurate.
However, for engineering components with hardening materials which are monotonically loaded, the proposed method offers further advantages:
\begin{itemize}
  \item significantly reduced computation time,
  \item hardening modeled by use of a plastic strain tensor as internal variable,
  \item any type of hardening behavior can be defined,
  \item real-world material parameters as needed for plasticity modeling can be used, and
  \item pure loading path is similar to classic plasticity.
\end{itemize}
This is also beneficial to engineering applications: known real-world material properties can be taken into account by our topology optimization at an early stage of research \& development.
Furthermore, optimization results including complex material models are available within a computation time which is appropriate for fast-moving development.

The paper is structured as follows: first, we recall the basics of the thermodynamic topology optimization by use of Hamilton's principle and complement our previous approaches by hardening including material behavior.
To this end, we develop a surrogate material model for our topology optimization approach that accounts for plastic strains without consideration of dissipation-related hysteresis effects.
Afterwards, we present a suitable strategy for numerical implementation.
Finally, the functionality of the proposed approach is tested and analyzed by means of computing topology optimizations for several boundary value problems.

\section{Thermodynamic topology optimization and a novel surrogate model for hardening materials}

The thermodynamic topology optimization is based on Hamilton's principle which is usually a variational strategy for material modeling \cite{junker2021extended}.
Assuming stationary of an extended Hamilton functional follows the thermodynamic and physical laws and yields field equations for all state variables, \ie{}, displacements, temperature and internal variables.
Expanding the functional for topology optimization provides the benefit that the optimization problem can be tackled by a system of partial differential equations.
Consequently, the mathematical optimization problem is converted into an evolutionary problem.
At the same time, the stationarity of the extended Hamilton functional comprises the evolution of microstructural material behavior which affects the evolution of topology.
Furthermore, constraints on the topology design as well as on the material behavior can be considered easily by taking use of Lagrange or Karush Kuhn Tucker parameters.
It is worth mentioning that no classical optimization problem is solved in thermodynamic topology optimization.
In contrast, the stationarity condition of the Hamilton functional with respect to the density variable serves as update scheme for the topology.

We use the following notation for tensor operations:
the single contraction is noted as ``$\cdot$'' reading $\bfa \cdot \bfb = c \Leftrightarrow a_i b_i = c$ when applied to two vectors $\bfa$ and $\bfb$, while it results in $\bfA \cdot \bfb = \bfc \Leftrightarrow A_{ij} b_j = c_i$ when applied to a vector $\bfb$ and a second-order tensor $\bfA$.
Moreover, the double contraction is denoted as ``$:$''.
It results in $\bfA : \bfB = c \Leftrightarrow A_{ij} B_{ij} = c$ when applied to two second-order tensors $\bfA$ and $\bfB$ while it results in $\dsA : \bfB = \bfC \Leftrightarrow \dsA_{ijkl} B_{kl} = C_{ij}$ when applied to a fourth-order tensor $\dsA$ and a second-order tensor $\bfB$.
Finally, the tensor product, \ie{}, the dyadic product, is noted as ``$\otimes$'' and reads $\bfa \otimes \bfb = \bfC \Leftrightarrow a_i b_j = C_{ij}$ when applied to two vectors $\bfa$ and $\bfb$ and $\bfA \otimes \bfB = \dsC \Leftrightarrow A_{ij} B_{kl} = \dsC_{ijkl}$ when applied to two second-order tensors $\bfA$ and $\bfB$.

In this contribution, the approach of topology optimization does not account for dynamic effects and therefore, we consider quasi-static loading.
Here, the extended Hamilton functional \cite{junker2021extended} for a quasi-static and isothermal case reads
\be
  \bar\calH = \bar\calH \left[\bfu,\bfy \right] := \calG\left[\bfu, \bfy \right] + \calD\left[\bfalpha \right]
\ee
and sums the Gibbs energy $\calG$ and the dissipation-related work $\calD$.
This functional depends on the displacements $\bfu$ and the state variable $\bfy = \{\bfalpha, \chi\}$.
The state variable is decomposed into the vectorial quantity $\bfalpha$ collecting all internal variables which describe the physical material behavior in terms of the microstructural state.
In our case of an hardening material, we thus chose $\bfalpha=\bfepsp$ where $\bfepsp$ denotes the plastic part of the strain tensor and $\bfeps^\mathrm{e}$ the elastic part, \ie{}, $\bfeps=\bfeps^\mathrm{e}+\bfepsp$.
The quantity $\chi$ denotes the density variable for defining the topology.
Here, the density variable $\chi \in [\chi_{\mathrm{min}},1]$ with $\chi_{\mathrm{min}} > 0$ represents void ``white'' material for $\chi = \chi_{\mathrm{min}}$, the full ``black'' material for $\chi = 1$, and a mixed ``gray'' phase for $\chi \in ]\chi_{\mathrm{min}},1[$.
The relative density is then modeled via the SIMP approach \cite{bendsoe2003Topology} by the interpolation function
\be
  \label{eq:chiInterpolation}
  \rho(\chi)=\chi^3 \, ,
\ee
for instance.
Other approaches are also possible, see~\cite{Junker2021hyperelastic} where a sigmoid function has been used.

According to Hamilton's principle the stationary condition of the functional is provided as
\be
 \bar\calH = \bar\calH[\bfu, \bfepsp, \chi] := \calG\left[\bfu, \bfepsp, \chi \right] + \calD\left[\bfepsp \right] \rightarrow \underset{\bfu,\bfepsp, \chi}{\text{stat}} \, .
\ee
Therein, $\calG$ is defined as difference between the energy stored in the body with volume $\Omega$ and the work due to external forces.
It hence reads
\be
  \label{eq:GibbsEnergy}
  \calG \left[\bfu, \bfepsp, \chi \right] := \int_{\Omega} \Psi(\bfeps^\mathrm{e}, \chi) \dd V - \int_{\Omega} \bfb^{\star} \cdot \bfu \dd V - \int_{\Gamma_{\sigma}} \bft^{\star} \cdot \bfu \dd A
\ee
with the Helmholtz free energy $\Psi$, the body forces $\bfb^{\star}$ and the traction vector $\bft^{\star}$.
The boundary conditions are defined as Dirichlet conditions for $\bfu^{\star}$ on $\Gamma_{u}$ and as Neumann conditions for $\bft^{\star}$ on $\Gamma_{\sigma}$.
Hence, the complete boundary $\partial\Omega$ of the body is given by $\partial\Omega = \Gamma_{u} \cup \Gamma_{\sigma}$ and $\Gamma_{u} \cap \Gamma_{\sigma} = \varnothing$.
Furthermore, the dissipation-related work is defined by
\be
  \calD\left[\bfepsp\right] := \int_{\Omega} \bfpdiss : \bfepsp \dd V
\ee
with the non-conservative force $\bfpdiss$ which can be derived from the dissipation function $\Delta^{\diss}$ by
\be
\label{eq:pdiss}
  \bfpdiss := \pf{\Delta^{\diss}}{\dot{\bfeps}^\mathrm{p}} \, .
\ee
More details on the thermodynamic basis are provided in~\cite{junker2021extended}.
According to \cite{Junker2021hyperelastic}, the physically motivated Hamilton functional $\bar{\calH}$ can be extended for thermodynamic topology optimization by adding
\be
  \label{eq:HamiltonFunctinalTTO}
  \calH\left[\bfu, \bfepsp, \chi \right] := \bar\calH \left[\bfu, \bfepsp, \chi \right] - \calR \left[\chi \right] + \calC \left[\bfepsp, \chi \right]
\ee
where additional constraints are included in $\calC$ and the rearrangement of topology is accounted for by the functional $\calR$, defined as
\be
  \calR \left[\chi \right] := \calD_{\chi} \left[\chi \right] + \calF \left[\chi \right] \, .
\ee
Here, the flux term 
\be
  \calF \left[\chi \right] := \int_{\Omega} \frac{1}{2} \, \beta \, \norm{\nabla \chi}^2 \dd V
\ee
accounts for the convective rearrangement with the regularization parameter $\beta > 0$.
It thus serves as gradient penalization for the density variable and also controls the members size via the parameter $\beta$.
Additionally, the source term
\be
  \calD_{\chi} \left[\chi \right] := \int_{\Omega} p^\diss_\chi \, \chi \dd V
\ee
accounts for local rearrangement.
Analogously to \eqref{eq:pdiss}, the non-conservative term for local rearrangement is assumed to be derivable from an associated dissipation function according to
\be
  p^\diss_\chi := \pf{\Delta^{\diss}_\chi}{\dot{\chi}} \, .
\ee
For the dissipation function, we follow \cite{jantos2019accurate} and chose
\be
  \Delta^{\diss}_\chi = \frac{1}{2} \, \eta\, \dot{\chi}^2 \, .
\ee
The viscosity parameter $\eta > 0$ controls the velocity of evolution of topology.
In this manner, the Hamilton functional \eqref{eq:HamiltonFunctinalTTO} is able to couple microstructure evolution and topology optimization.
We propose that an optimal structure can be found if this functional becomes stationary.

The stationary condition with respect to all variables 
\be
  \delta\calH = \delta_{\bfu} \calH + \delta_{\bfepsp} \calH + \delta_{\chi} \calH = 0 \qquad \forall\; \delta\bfu, \delta\bfepsp, \delta\chi
\ee
yields the following system of governing equations
\be
  \label{eq:generalFieldEquations}
  \begin{cases}
    \D \delta_{\bfu} \calH = 0 = \int_{\Omega} \pf{\Psi}{\bfeps} : \delta\bfeps \dd V - \int_{\Omega} \bfb^{\star} \cdot \delta\bfu \dd V - \int_{\Gamma_{\sigma}} \bft^{\star} \cdot \delta\bfu \dd A & \forall\; \delta\bfu \\[4mm]
    \D \delta_{\bfepsp} \calH = 0 = \int_{\Omega} \pf{\Psi}{\bfepsp} : \delta\bfepsp \dd V + \int_{\Omega} \pf{\Delta^{\diss}}{\dot{\bfeps}^\mathrm{p}} : \delta\bfepsp \dd V + \delta_{\bfepsp} \, \calC & \forall\; \delta\bfepsp \\[4mm]
    \D \delta_{\chi} \calH = 0 = \int_{\Omega} \pf{\Psi}{\chi} \ \delta\chi \dd V - \int_{\Omega} \eta \, \dot\chi \, \delta\chi \dd V - \int_{\Omega} \beta \ \nabla \chi \cdot \nabla\delta\chi \dd V + \delta_{\chi} \, \calC & \forall\; \delta\chi
  \end{cases}
\ee
where each equation belongs to one of the independent system variables, cf. also~\cite{Junker2021hyperelastic} for a general microstructure consideration in case of finite deformations.
Here, the standard notation $\delta\bfeps:=\mathrm{sym}(\nabla\otimes\delta\bfu)$ is used.
The first condition is identified as the weak form of the balance of linear momentum where the stress is given by $\bfsigma = \partial\Psi / \partial\bfeps$.
The second condition constitutes as governing equation for the plastic strains $\bfepsp$ and the last equation is the field equation for topology optimization.

\subsection{Specification of the energetic quantities and the constraints}
\label{sec:specificationDerivation}
The system of governing equations \eqref{eq:generalFieldEquations} establishes the general framework for the optimization process.
However, by specification of the free energy density $\Psi$, the dissipation function $\Delta^{\diss}$, and the constraint functional $\calC$ the characteristics of the model for the behavior of hardening materials and the density variable are defined.
To this end, the objective and constraints corresponding to conventional optimization algorithms are included directly into derivation here.

For the free energy, we follow the classic approach of elasto-plastic materials and combine it with the relative density $\rho(\chi)$ in \eqref{eq:chiInterpolation}.
This gives
\be
  \Psi(\bfeps^\mathrm{e},\chi) = \rho(\chi) \, \Psi_0  := \frac{1}{2} \, (\bfeps-\bfepsp): \dsEChi : (\bfeps-\bfepsp)
\ee
where the stiffness tensor of the full material is given by $\dsE_0$ and the energy of the virtually full material is given by
\be
  \label{eq:fullHelmholtzEnergy}
  \Psi_0 := \frac{1}{2} \, (\bfeps-\bfepsp): \dsE_0 : (\bfeps-\bfepsp) \, .
\ee
Consequently, we obtain for the stresses
\be
  \label{eq:stress}
  \bfsigma = \pf{\Psi}{\bfeps}= \dsEChi : (\bfeps-\bfepsp) \, .
\ee
The derivative of $\Psi$ with respect to $\bfepsp$ thus yields
\be
  \pf{\Psi}{\bfepsp} = - \bfsigma
\ee
and the derivative of $\Psi$ with respect to $\chi$ yields
\be
  \label{eq:DFTopOpt}
  \drivingForce:= -\pf{\Psi}{\chi} = - 3 \, \chi^2 \, \Psi_0
\ee
as driving force or sensitivity for the topology optimization, respectively.
The driving force~$\drivingForce$ is non-zero for all conditions with $\Psi_0\not=0$ since $\chi\geq\chi_\text{min}$.
Furthermore, the evolution of plastic strains influences $\Psi_0$ and, in turn, the driving force $\drivingForce$ and thus the update condition for optimization, cf.~\eqref{eq:generalFieldEquations}$_3$.

For modeling the stress/strain relationship of hardening materials we make use of the following assumption:
we mentioned the ``virtual'' increase and decrease of plastic strains during an evolutionary optimization process.
This means, differently stiff structures loaded with the same external loading result in different displacement fields.
For instance, high strains might be present in the beginning of the optimization process with associated high plastic strains.
However, the evolution of local stiffness results in reduced strains and consequently reduced plastic strains which we denote as ``virtual unloading''.
In a classic elasto-plastic material model, the virtual unloading evokes dissipation
\be
  \label{eq:dissfunc}
  \Delta^{\diss} = r \, \norm{\dot{\bfeps}^\mathrm{p}}
\ee
with the yield limit $r$ which results in the typical hysteresis curve.
This approach yields a rate-independent formulation; details can be found, \eg{}, in \cite{junker2021extended}.
However, we are interested in a material model that computes a single loading point on the stress/strain diagram for each displacement state as it results from physical loading.
Therefore, in the case of virtual unloading, the loading branch in the stress/strain curve needs to be followed back.
To this end, we propose a hysteresis-free behavior by postulating a vanishing dissipation function, \ie{},
\be
  \Delta^{\diss}=0 \, .
\ee
For this reason, the surrogate model is not path-dependent as classic plasticity models.

Moreover, the definition of the yield criterion is omitted without dissipation and therefore, the yield condition is included as a constraint by demanding
\be
  \norm{\bfsigmadev} = r
\ee
during plastic evolution where the stress deviator $\bfsigmadev= \bfsigma - 1/3 \ \tr{\bfsigma} \bfI$ is computed by
\be
  \bfsigmadev = \dsP : \bfsigma
\ee
with the projection tensor $\dsP$.
The threshold value $r$ will be defined phenomonologically and needs to be combined with the relative density $\rho\left(\chi\right)$ according to \cite{Duysinx1998Topology} for physical consistency.
It is thus possible to formulate any type of hardening without adapting the surrogate model itself.
A special case is non-hardening which results as stress plateau corresponding to ideal plasticity as special case of plasticity.
Non-hardening behavior is determined by a constant material parameter, \eg{}, the yield stresses $\sigmaY$, which yields
\be
 \label{eq:r_ideal}
  r = \chi^3 \, \sigmaY \, .
\ee
Any other type of hardening can be described by choosing a non-constant $r = r(\norm{\bfepsp})$.
To this end, we propose linear hardening by defining
\be
  \label{eq:r_linHard}
  r = r(\norm{\bfepsp}) = \chi^3 \left( \sigmaY + h \, \norm{\bfepsp} \right)
\ee
with the slope of hardening curve $h$ and exponential hardening according to \cite{junker2017numerical} by
\be
  \label{eq:r_expHard}
  r = r(\norm{\bfepsp}) = \chi^3 \left( \sigmaY + h_{1} \, \norm{\bfepsp} + \frac{1}{\kappa} \, \left( h_{1} - h_{0} \right) \left( e^{-\kappa \, \norm{\bfepsp}} - 1 \right) \right) \, .
\ee
Here, $h_0$ denotes the initial and $h_1$ the end slope of the hardening curve and $\kappa$ controls the transition from $h_0$ to $h_1$.
Since our approach is equivalent for different definitions of $r$, it is possible to best represent a real material behavior depending on standard material parameters.
We always use the general notation $r$ as yield criterion in the following.

Lastly, a constraint for the surrogate model is that the hydrostatic stress based only on the strains $\bfeps$ and reads
\be
  \bfsigma^{\mathrm{h}} = \bfI : \bfsigma = \bfI : \dsEChi : \bfeps \, .
\ee
Consequently, plastic strains are volume-preserving which is an experimentally known characteristic of plastic deformations.

The limitation of the stress norm by the yield threshold and the reformulated hydrostatic stress are included through the constraint functional by using the Lagrange parameters $\lambda_\mathrm{\sigma}$ and $\lambda_\mathrm{V}$, respectively.

It remains to identify the constraints for the density variable $\chi$ to finally formulate the constraint functional $\calC$.
The first constraint is given by the interval in which $\chi$ is defined: values of $\chi$ that are negative are not reasonable; same limitation holds true for values of $\chi$ that are larger than one.
Consequently, we demand $\chi \in [\chi_{\mathrm{min}},1]$ where the lower bound is set to a small value $1 \gg \chi_{\mathrm{min}} > 0$ due to numerical reasons.
These bounds are taken into account by use of a Karush Kuhn Tucker parameter $\gamma$.
Furthermore, the volume of the topology relative to the total design volume is prescribed by the parameter $\relStructVol$.
Consequently, it has to hold
\be
  \int_\Omega \chi \dd V = \relStructVol \, \Omega
\ee
which is included to the constraint functional by use of a Lagrange parameter $\lambda_\mathrm{\chi}$.

Combining these four constraints, \ie{}, norm of the stress deviator being equivalent to the yield threshold $r$, the restated hydrostatic stress $\bfsigma^{\mathrm{h}}$, bounded interval for $\chi$, and control of the total relative structure volume $\relStructVol$, we finally arrive at
\begin{multline}
  \label{eq:ConstraintsFunctional}
  \calC\left[\bfepsp, \chi \right] := \lambda_\mathrm{\sigma} \int_{\Omega} \left(\norm{\bfsigmadev} - r\right) \dd V   + \lambda_\mathrm{V} \int_{\Omega} \left( \bfI : \bfsigma - \bfI : \dsEChi : \bfeps \right) \dd V \\
  + \int_{\Omega} \gamma \, \chi \dd V + \lambda_\mathrm{\chi} \left(\int_{\Omega} \chi \dd V - \relStructVol \, \Omega \right)  \, .
\end{multline}

\subsection{The stationarity condition with respect to the plastic strains}
\label{sec:stationarityConditionPlasticStrains}
It remains to appropriately analyze the stationarity condition of the Hamilton functional with respect to the plastic strains.
This condition enables us to compute the plastic strains which, in combination with the total strain, specify the stress state.
To this end, we use the specifications for a vanishing dissipation function $\Delta^{\diss}$ and the constraint functional \eqref{eq:ConstraintsFunctional} to evaluate \eqref{eq:generalFieldEquations}$_2$ as
\bea
  \label{eq:StatPlastLoc}
  \int_\Omega \left( -\bfsigma + \lambda_\mathrm{\sigma} \left[ \pf{\norm{\bfsigmadev}}{\bfepsp} - \dr \right] - \lambda_\mathrm{V} \, \bfI : \dsEChi \right) :\delta\bfepsp \dd V &=& 0 \qquad \forall \; \delta \bfepsp \notag \\
  \Rightarrow\quad - \bfsigma + \lambda_\mathrm{\sigma} \left[ \pf{\norm{\bfsigmadev}}{\bfepsp} - \dr \right] - \lambda_\mathrm{V} \, \bfI : \dsEChi &=& \boldsymbol{0} \notag\\
  \Leftrightarrow - \bfsigma -\lambda_\mathrm{\sigma} \left[ \frac{\bfsigmadev}{\norm{\bfsigmadev}} :\dsP: \dsEChi + \dr \right] - \lambda_\mathrm{V} \, \bfI : \dsEChi &=& \boldsymbol{0} \, .
\eea
Solving \eqref{eq:StatPlastLoc} for the plastic strains constitutes our surrogate model for hardening material behavior.
A detailed derivation of the Lagrange multipliers is deferred to App.~\ref{appendix:DerivationSurrogateModel}.
There, we show that the governing equation for the plastic strains is given as
\bea
  \label{eq:s_epsP}
  \bfs &:=& - \bfsigma + \frac{r^3}{\D \bfsigmadev :  \dsEChi : \bfsigmadev + \pf{r}{\bfepsp} : \bfsigmadev \, r} \, \left[\frac{\bfsigmadev:\dsEChi}{r} + \pf{r}{\bfepsp} \right] \\
  && + \frac{\bfI : \dsEChi : \bfeps}{\bfI : \dsE_0 : \bfI} \, \bfI : \dsE_0 = \boldsymbol{0} \notag
\eea
which is a non-linear algebraic equation.
The derivative of the yield criterion $r$ is defined as
\be
  \label{eq:dr}
  \D \pf{r}{\bfepsp} = \begin{cases}
  \D \boldsymbol{0} & \text{non-hardening} \\
  \D \pf{\rH}{\bfepsp} = \pf{\rH}{\norm{\bfepsp}} \, \pf{\norm{\bfepsp}}{\bfepsp} = r' \, \frac{\bfepsp}{\norm{\bfepsp}} & \text{hardening}
\end{cases}
\ee
where the term $r'$ for the defined types of hardenings reads
\be
  \D r' = \begin{cases}
  \D h & \text{linear hardening} \\
  \D  h_1 - \left( h_{1} - h_{0} \right) \, e^{-\kappa \, \norm{\bfepsp}} & \text{exponential hardening}
\end{cases} \, .
\ee
In case of non-hardening with $r = \text{constant}$ and the derivative from \eqref{eq:dr}, we can reduce \eqref{eq:s_epsP} to
\be
  \label{eq:s_epsP_ideal}
  \bfs^{\mathrm{nh}} := - \bfsigma + \frac{r^2}{\bfsigmadev :  \dsE_0 : \bfsigmadev} \, \bfsigmadev:\dsE_0 + \frac{\bfI : \dsEChi : \bfeps}{\bfI : \dsE_0 : \bfI} \, \bfI : \dsE_0 = \boldsymbol{0} \, .
\ee

\textit{Remark:} it is worth mentioning that we do not receive a differential equation for the internal variable as it is usually the case for classic elasto-plastic models.
This routes back to assuming a dissipation-free evolution of the plastic strains which, in turn, are determined by energy minimization.

Components of the plastic strain tensor only evolve to compensate high stresses which are greater than the yield stress $\sigmaY$.
Therefore, it is mandatory to identify a suitable criterion for distinguishing whether an elastic or hardening material behavior is present.
Since the purpose of the modified surrogate model is to display the same material behavior for loading like a classic material model for elasto-plasticity, we make use of the indicator function that would result from the dissipation function in \eqref{eq:dissfunc} via a Legendre transformation, cf. \cite{junker2021extended}.
This indicator function reads
\be
  \label{eq:YieldFunctionSigma}
  \Phi_{\sigma} = \norm{\bfsigmadev} - r \le 0
\ee
where elastic behavior is present for $\Phi_{\sigma} <0$ and plastic / hardening behavior for $\Phi_{\sigma} =0$.

Fitting the characteristics of the classic elasto-plastic material model, physical unloading from a plastic state can be detected by this indicator function when the stress decreases once again below the yield threshold $r$.
The elastically stored energy is released first and the residual, plastic strains remains.
In this way, the hysteresis loop in the stress/strain diagram of a physical material evolves.

This behavior is suppressed by the surrogate material model as discussed above.
Virtual unloading from a plastic state should immediately result in a decrease of plastic strains.
Thus, the plastic strains are reduced first and only if no plastic strains are present anymore, the elastically stored energy is released.
In this way, the loading branch in the stress/strain curve is followed both for loading and virtual unloading.

Consequently, the stress is not a suitable measure for the indicator function related to the surrogate model.
In contrast, the strains are identified as suitable measure.
We therefore reformulate the indicator function \eqref{eq:YieldFunctionSigma} in terms of strains.
To this end, the yield threshold $r$ is compared to the stress $\bfsigma^{\star} = \dsEChi : \bfeps$ which occurs depending on the total strain $\bfeps$.
Therefore, we can present the yield function as
\be
  \label{eq:YieldFunctionPlasticStrains}
  \Phi_{\varepsilon} = \norm{\dsP : \bfsigma^{\star}} - r \qquad \text{where } \qquad
  \Phi_{\varepsilon} = \begin{cases}
    < 0    & \text{elastic} \\
  \geq 0 & \text{hardening}
  \end{cases} \, .
\ee
This indicator function is similar to that one used for Hencky plasticity with $r = \sigmaY$, cf. \cite{Duester2001ThePVersion}, which confirms that both models are only feasible for monotonic loading.

\subsection{The stationarity condition with respect to the density variable}
\label{sec:derivationStationarityDensityVariable}
Finally, the evolution of the density variable needs to be formulated.
Therefore, it remains to investigate the governing equation for the density variable $\chi$ which is given by \eqref{eq:generalFieldEquations}$_3$.
Making use of the constraint functional $\calC$ in \eqref{eq:ConstraintsFunctional} and the driving force for topology optimization $\drivingForce$ in~\eqref{eq:DFTopOpt}, the stationarity with respect to $\chi$ takes the form
\be
  \label{eq:TopOptWeak}
  \int_\Omega \left( -\drivingForce - \eta \, \dot{\chi} + \gamma + \lambda_\mathrm{\chi} \right) \delta\chi \dd V - \int_\Omega \beta \, \nabla\chi \cdot\nabla\delta\chi \dd V = 0 \qquad \forall \; \delta\chi
\ee
which is a parabolic differential equation and shows some similarities to phase field equations, cf.~\cite{bartels2021cahn} for instance.
Analogously to the stationarity with respect to the displacements in \eqref{eq:generalFieldEquations}$_1$, this equation \eqref{eq:TopOptWeak} is the weak form of the associated Euler equation (which is the balance of linear momentum for the displacements).
Therefore, one possibility for numerical evaluation would be given by direct application of the finite element method.
A comparable approach has been presented in \cite{junker2015variational}.
However, it has turned out that this procedure is much more time-consuming than applying the numerical method that has been presented in \cite{jantos2019accurate} due to the complex constraints of the bounded interval for $\chi$ and the prescribed total density $\relStructVol$.
Therefore, in order to apply the method of the previous work in \cite{jantos2019accurate} which reduces the numerical efforts by approximately one order of magnitude, we transform \eqref{eq:TopOptWeak} to its strong form by integration by parts.
This results in 
\be
  \label{eq:TopOptStrong}
  \begin{cases}
    \D \eta \, \dot\chi \, \in - \drivingForce + \beta \, \lap\chi + \lambda_\mathrm{\chi} + \gamma & \forall\; \bfx \in \Omega \\[4mm]
    \D \bfn \cdot \nabla\chi = 0 & \forall\; \bfx \in \partial\Omega
  \end{cases}
\ee
where \eqref{eq:TopOptStrong}$_2$ is the Neumann boundary condition for the density variable.
It ensures conservation of the prescribed total structure volume.
Meanwhile, the change of the density variable is defined by \eqref{eq:TopOptStrong}$_1$ and accounts for the Laplace operator which is defined as
\be
  \lap\chi:= \frac{\partial^2 \chi}{\partial x^2} + \frac{\partial^2 \chi}{\partial y^2} + \frac{\partial^2 \chi}{\partial z^2} \, .
\ee
The transient characteristic of this term require the specification of an initial value for $\chi(\bfx,t=0)=\chi_\mathrm{ini}\,\forall \;\bfx\in\Omega$, which will be introduced with the numerical treatment in \secref{sec:updateDensity}.

\FloatBarrier
\section{Numerical implementation}
\label{sec:numerics}
In summary, the following system of coupled differential-algebraic equations needs to be solved:
\be
\label{eq:numericalFieldEquations}
\begin{cases}
  \D 0 = \int_\Omega \bfsigma \cdot \delta\bfeps \dd V - \int_\Omega \bfb^* \cdot \delta\bfu \dd V - \int_{\partial\Omega} \bft^* \cdot \delta\bfu \dd A & \forall \; \delta\bfu \\[4mm]
  \D \boldface{0}  = \bfs \text{, see \eqref{eq:s_epsP}} & \forall \; \bfx \in \Omega\\[4mm]
  \D \dot{\chi} \, \in \, \frac{1}{\eta} \left( - \drivingForce + \beta \, \lap\chi + \lambda_\mathrm{\chi} + \gamma \right) & \forall \; \bfx \in \Omega
\end{cases}
\ee
The numerical implementation based on this solution is written in Julia programming language \cite{Julia} and published as open-access file in \cite{Kick2022Implementation}.
It is worth mentioning that we use for now on the usual Voigt notation for the stresses and strains which reduces, for instance, the double contraction to a scalar product in \eqref{eq:numericalFieldEquations}$_1$ and \eqref{eq:numericalFieldEquations}$_2$.

The numerical solution of the system of equations of the displacement field $\bfu$, the microstructural plastic strains $\bfepsp$ and the topology density $\chi$ is a sophisticated task due to the inherent non-linearities, constraints, and strong coupling.
Therefore, we solve equation \eqref{eq:numericalFieldEquations}$_1$ by finite element method (FEM) where \eqref{eq:numericalFieldEquations}$_2$ is solved on each integration point to capture a physically correct microstructure behavior.
This non-linear computation leads to a non-linear FEM problem.
In addition, \eqref{eq:numericalFieldEquations}$_3$ is solved by the finite difference method (FDM).
Hence, both the FEM, in a monolithic manner, and the FDM are employed for the solution. 
This combination in a staggered process is referred to as neighbored element method (NEM), cf.~\cite{jantos2019accurate}.
The staggered process can be interpreted as operator split which has turned beneficial in our previous works as in~\cite{jantos2019accurate} and also for adaptive finite element usage in~\cite{vogel2020adaptive}.
According to the staggered process, our method shows similarities to conventional mathematical optimization methods which are composed of alternating structure computation and optimization algorithm.

During the iterative solution of \eqref{eq:numericalFieldEquations}, each iteration step corresponds to an update step of the thermodynamic topology optimization.
In this way, an evolutionary update of, \eg{}, the density field takes place.
For this purpose, we employ a standard discretization in pseudo-time, given as
\be
  \Delta t := t_{n+1} - t_n 
\ee
where $t_{n+1}$ refers to the current iteration step and $t_n$ to the previous iteration step.

\subsection{Update of the displacements}
A standard non-linear finite element approach is employed for updating the displacements and the stress in~\eqref{eq:numericalFieldEquations}$_1$ is consistently evaluated as
\be
  \label{eq:numStress}
  \bfsigma_{n+1} = \chi^3_n \, \dsE_0 \cdot(\bfeps_{n+1} - \bfepsp_{n+1}) 
\ee
for the current time step $n+1$.
For the FEM discretization, the displacement field is approximated using the Galerkin-Ansatz
\be
  \label{eq:GalerkinAnsatz}
  u_{k} = N_{o} \, u_{o}^\comp{k} = \bfN \cdot \hat\bfu^\comp{k}
\ee
with the shape function $\bfN$ and the nodal displacement $\hat\bfu^\comp{k}$ in the spatial direction $k$.
Therefore, the weak form of the balance of linear momentum in~\eqref{eq:numericalFieldEquations}$_1$ transforms to
\be
  \label{eq:updateDisplacements}
  \int_\Omega \bfsigma \cdot \delta\bfeps \dd V = \delta\hat\bfu \cdot \int_\Omega \bfB^\mathrm{T} \cdot \bfsigma \dd V = 0 =: \delta\hat\bfu \cdot \hat\bfr \qquad \forall\; \delta\hat\bfu
\ee
when body forces are neglected.
Here, $\bfB$ denotes the usual operator matrix including the spatial derivatives of the shape function.
The quantity $\delta\hat\bfu$ is the global column matrix of nodal virtual displacements which also includes the Dirichlet boundary conditions.
Finally, the global residual column matrix is denoted by $\hat\bfr$ and, accordingly, the nodal displacements will be found from $\hat\bfr = \boldsymbol{0}$.
The global residual $\hat\bfr$ is assembled in usual manner by
\be
  \hat\bfr := \AssemargPJ{e}{} \hat\bfr_e \quad\text{where}\quad \hat\bfr_e := \int_{\Omega_e} \bfB^\mathrm{T}_e \cdot \bfsigma \dd V
\ee
denotes the residual column matrix for each element $e$.
More details on the finite element method can be found in standard textbooks, \eg{}, \cite{wriggers2008nonlinear}.

In our numerical implementation (cf. \cite{Kick2022Implementation}) of the thermodynamic topology optimization including hardening materials, we made use of the finite element toolbox Ferrite.jl~\cite{Ferrite}.
Ferrite.jl uses a gradient-based equation solver as it is the standard for many finite element programs.
Consequently, the iterative solution  process for  $\hat\bfr=\boldsymbol{0}$ is performed by
\be
  \hat\bfr_{j+1} = \hat\bfr_{j} + \pf{\hat\bfr}{\hat\bfu} \cdot \Delta\hat\bfu = \boldsymbol{0}
\ee
where the iteration number is given by $j$.
The increment $\Delta\hat\bfu$ updates the displacement field iteratively for the current plastic strains $\bfepsp_{j+1}$ and the fixed density field $\chi_n$ by
\be
  \hat\bfu_{j+1} = \hat\bfu_{j} - d \, \left(\bfK_{j}^{-1} \cdot \hat\bfr_{j} \right) \, .
\ee
Here, the update is damped by the parameter $d$.
Consequently, the current residual $\hat\bfr_{j+1}$ depends on the nodal displacements $\hat\bfu_{j}$ and the damping parameter $d$ as a function $\hat\bfr_{j+1} := \hat\bfr\left(\hat\bfu_{j+1}\right) = \hat\bfr\left(\hat\bfu_{j},d\right)$.
The required element tangent is computed as
\be
  \bfK_{e} = \pf{\hat\bfr_{e}}{\hat\bfu_{e}} = \int_{\Omega_{e}} \bfB_{e}^\mathrm{T} \cdot \chi_n^3 \, \dsD_0 \cdot \bfB_{e} \dd V
\ee
with the column matrix of displacements for each finite element $e$ denoted as $\hat\bfu_{e}$.
The consistent tangent operator $\dsD_0 = \mathrm{d}\bfsigma_{j+1} / \mathrm{d}\bfeps_{j+1}$ with $\bfsigma_{j+1} := \bfsigma\left(\bfeps_{j+1},\bfepsp_{j+1},\chi_{n}\right)$ depends on the updated plastic strains $\bfepsp_{j+1}$ and therefore will be defined as \eqref{eq:consistentTangentOperator} in \secref{sec:UpdateEpsP}.
Then, the assembled tangent is constructed by
\be
  \bfK = \pf{\hat\bfr}{\hat\bfu} = \AssemargPJ{e}{} \pf{\hat\bfr_{e}}{\hat\bfu_{e}}
\ee
where $\bfK = \bfK\left(\hat\bfu\right)$ so that the current tangent can be defined as $\bfK_{j+1} := \bfK\left(\hat\bfu_{j+1}\right) = \bfK\left(\hat\bfu_{j},d\right)$.

Computing a solution for the non-linear material behavior is challenging due to the possibly remarkable rearrangement of the topology.
Therefore, numerical damping of the Newton iterations is beneficial for a smooth convergence.
However, damping is costly and should be used only when necessary.
Therefore, damping is deactived when the Newton iterations of the FEM simulation of the previous global iteration are lower or equal to three which proposes a rapid convergence in the current simulation as well.
Otherwise, damping is activated if the number of current iterations takes more than two additional steps than in the previous load step.

In case of damping, the damping parameter $d$ is set in an adaptive manner.
Hence, we define the energy function of the system according to \cite{wriggers2008nonlinear} as
\be
  g\left(d\right) = - \hat\bfr\left(\hat\bfu_{j},0\right) \cdot \hat\bfr\left(\hat\bfu_{j},d\right) = 0
\ee
which is solved for the damping parameter $d$.
The computation of the damping parameter requires a sign change of the current energy~$g\left(0\right)$ and the forward-looking energy~$g\left(1\right)$ so that $d \in ]0,1]$.
In addition, it is only activated if $g\left(1\right) > 0.3 \, g\left(0\right)$.
Then, the numerical damping parameter $d$ will be computed once by
\be
  d = 1 - \frac{\hat\bfr\left(\hat\bfu_{j},0\right) \cdot \hat\bfr\left(\hat\bfu_{j},0\right)}{\hat\bfr\left(\hat\bfu_{j},0\right) \cdot \bfK\left(\hat\bfu_{j},1\right) \cdot \bfK^{-1}\left(\hat\bfu_{j},0\right) \cdot \hat\bfr\left(\hat\bfu_{j},0\right)}
\ee
and is restricted by $d = \max\{d_{\min},\min\{d,1\}\}$ with $d_{\min} = 0.1$.
Otherwise, if no damping is necessary, an update with full step size can be run with $d = 1$.

\textit{Remark:} It is worth mentioning that we used the package Tensors.jl \cite{JuliaTensors} in our numerical implementation which is optimized for using tensors of higher order.
Therefore, we did not perform a finite element programming in standard form, \ie{}, by using the Voigt notation, but used the full tensor notation.
This, of course, also effects the dimensions of the other quantities, \ie{}, the $\bfB$ operator is an array with three indices.
For a more usual presentation, we presented the formulas by using the Voigt notation and deferred our array-based programming using the tensors package to \appref{appendix:FerriteFEM}.

\subsection{Update of the plastic strains}
\label{sec:UpdateEpsP}
The plastic strains are defined, as usual, for each integration point.
According to the discretization we employ for the density variable, all integration points in the same finite element are evaluated with the same value for the density variable $\chi$.
More details are given in \secref{sec:updateDensity} when we discuss the numerical treatment for the density variable.

The plastic strains are determined from solving~\eqref{eq:numericalFieldEquations}$_2$ which is a non-linear algebraic equation.
Within the update scheme of the plastic strains, we employ the operator split with $\bfsigma={\bfsigma}(\bfeps_{j+1},\bfepsp_{j+1},\chi_n)$ accounting for the element-wise density from the last iteration $n$ and the updated value of the plastic strains.
For the numerical implementation, we make use of Newton's method to find the roots of $\bfs$ and define the Newton iterator $i$.
Newton's method for~\eqref{eq:s_epsP} reads
\be
  \bfs + \pf{\bfs}{\bfepsp} \cdot \Delta \bfepsp = \boldsymbol{0}
\ee
and the plastic strains are iteratively updated according to
\be
  \label{eq:NewtonUpdateESepsp}
  \bfepsp_{j+1} \leftarrow \bfepsp_{i+1} = \bfepsp_{i} - \left[ \pf{\bfs_{i}}{\bfepsp_{i}} \right]^{-1} \cdot \bfs_{i} \, .
\ee
Hence, the current plastic strains $\bfepsp_{j+1}$ for the FEM iteration $j+1$ correspond to the converged update.
The tangent reads
\bea
  \label{eq:ds_epsP}
  \dsdepsp &=& \dsEChi + \frac{1}{\left(\D \bfsigmadev \cdot \dsEChi \cdot \bfsigmadev + \dr \cdot \bfsigmadev \, r \right) ^2}  \\
  && \Bigg[ 3 \, r^2 \, \dr \left( \bfsigmadev \cdot \dsEChi \cdot \bfsigmadev + \dr \cdot \bfsigmadev \, r \right) \notag \\
  && - r^3 \bigg( - 2 \, \dsEChi \cdot \dsEChi \cdot \bfsigmadev + \ddr \cdot \bfsigmadev \, r \notag \\
  && - \dr \cdot \dsEChi \, r + \dr \cdot \bfsigmadev \cdot \dr \bigg) \Bigg] \otimes \left( \frac{1}{r} \, \bfsigmadev \cdot \dsEChi + \dr \right) \notag \\
  && + \frac{r^3}{\D \bfsigmadev \cdot \dsEChi \cdot \bfsigmadev + \dr \cdot \bfsigmadev \, r} \notag \\
  && \left[- \frac{1}{r^2} \, \left( \dsP \cdot \dsEChi \cdot \dsEChi \, r + \bfsigmadev \cdot \dsEChi \otimes \dr \right) + \ddr \right] \notag
\eea
where $\dsdepsp = \dsdepsp \left(\bfeps,\bfepsp,\chi\right) := \partial \bfs / \partial \bfepsp$.
Consequently, $\partial \bfs_{i} / \partial \bfepsp_{i} = \dsdepsp \left(\bfeps_{j+1},\bfepsp_{i},\chi_{n}\right)$.
The yield criterion $r$ was defined in \eqref{eq:r_linHard} and \eqref{eq:r_expHard} as well as its first derivative in \eqref{eq:dr}.
The second derivative of the yield criterion $r$ reads
\be
  \ddr = \begin{cases}
    \D \boldsymbol{0} & \text{non-hardening} \\
    \D \ppf{\rH}{\bfepsp} = r' \, \ppf{\norm{\bfepsp}}{\bfepsp} + \pf{r'\left(\norm{\bfepsp}\right)}{\norm{\bfepsp}} \, \pf{\norm{\bfepsp}}{\bfepsp} \otimes \frac{\bfepsp}{\norm{\bfepsp}} \\
    \D \hphantom{\ppf{\rH}{\bfepsp}} = r' \, \ppf{\norm{\bfepsp}}{\bfepsp} + r'' \, \frac{\bfepsp \otimes \bfepsp}{\norm{\bfepsp}^2} & \text{hardening}
  \end{cases}
\ee
where we make use of
\be
  \ppf{\norm{\bfepsp}}{\bfepsp} = \frac{\dsI}{\norm{\bfepsp}} - \frac{\bfepsp \otimes \bfepsp}{\norm{\bfepsp}^3} \, .
\ee
Furthermore, $r''$ is defined in terms of the type of the hardening as
\be
  \D r'' = \begin{cases}
  \D 0 & \text{linear hardening} \\
  \D  \kappa \, \left( h_{1} - h_{0} \right) \, e^{-\kappa \, \norm{\bfepsp}} & \text{exponential hardening}
\end{cases} \, .
\ee
The initial value for the plastic strains is chosen as $\bfepsp_{\mathrm{ini}}=\boldsymbol{0}$ at the beginning of each update.
The convergence is defined such that all components of $\bfs$ must be numerically zero, $\max\{\bfs\} \leq 10^{-8}$ for instance.

It turns out that the components of $\bfs$ are small for each integration point located at every element with a small density variable $\chi$.
For this reason, the value of plastic strains computed by the described method are not as accurate as for larger density values.
Therefore, we propose to factorize equation \eqref{eq:s_epsP} with $1 / \chi_{n}$ so that it reads
\be
  \label{eq:s_epsP_scaled}
  \tilde\bfs := \frac{1}{\chi_{n}} \, \bfs  = \boldsymbol{0}
\ee
and its tangent \eqref{eq:ds_epsP} can be denoted as
\be
  \label{eq:ds_epsP_scaled}
  \pf{\tilde\bfs}{\bfepsp} = \frac{1}{\chi_{n}} \, \pf{\bfs}{\bfepsp} \, .
\ee
Obviously, the scaling is only a numerical technique which has no influence on the magnitude of the resulting value but on the precision.
An overview of this numerical update algorithm is given in \algoref{alg:updatePlasticStrain}.
\begin{algorithm}[!htb]
  \caption{Compute update of plastic strains by Newton's method}
  \label{alg:updatePlasticStrain}
  \input{./03_figures/alg_update_epsP.tex}
\end{algorithm}

It is worth noting that due to the monolithic FEM the current consistent tangent operator
\be
  \label{eq:consistentTangentOperator}
  \dsD_0 = \tf{\bfsigma_{j+1}}{\bfeps_{j+1}} = \pf{\bfsigma_{j+1}}{\bfeps_{j+1}} + \pf{\bfsigma_{j+1}}{\bfepsp_{j+1}} \cdot \tf{\bfepsp_{j+1}}{\bfeps_{j+1}} = \dsE_0 - \dsE_0 \cdot \tf{\bfepsp_{j+1}}{\bfeps_{j+1}}
\ee
does depend on the current plastic strains $\bfepsp_{j+1}$ and is therefore not constant in each FEM iteration.
The plastic strains $\bfepsp_{j+1}$ depend on the total strains $\bfeps_{j+1}$ in accordance with the surrogate material model.
Therefore, the derivative
\be
  \label{eq:femTangentMaterial}
  \tf{\bfepsp_{j+1}}{\bfeps_{j+1}} = - \left[\pf{\bfs_{j+1}}{\bfepsp_{j+1}} \right]^{-1} \cdot \pf{\bfs_{j+1}}{\bfeps_{j+1}}
\ee
is given by $\bfs = \boldsymbol{0}$ with \eqref{eq:s_epsP} as implicit function.
Here, the first derivative in \eqref{eq:femTangentMaterial} is already given with $\partial \bfs_{j+1} / \partial \bfepsp_{j+1} = \bfP \left(\bfeps_{j+1},\bfepsp_{j+1},\chi_{n} \right)$ in \eqref{eq:ds_epsP}.
The second derivative is computed to be
\bea
  \label{eq:ds_eps}
  \dsdeps &=& - \dsEChi - \frac{r^3}{\left(\D \bfsigmadev \cdot \dsEChi \cdot \bfsigmadev + \dr \cdot \bfsigmadev \, r \right) ^2} \\
  && \left( 2 \, \dsEChi \cdot \dsEChi \cdot \bfsigmadev + \dr \cdot \dsEChi \, r \right) \otimes \left( \frac{1}{r} \, \bfsigmadev \cdot \dsEChi + \dr \right) \notag \\
  && + \frac{r^3}{\D \bfsigmadev \cdot \dsEChi \cdot \bfsigmadev + \dr \cdot \bfsigmadev \, r} \, \frac{1}{r} \, \dsP \cdot \dsEChi \cdot \dsEChi + \frac{\bfI \cdot \dsEChi}{\bfI \cdot \dsE_0 \cdot \bfI} \otimes \bfI \cdot \dsE_0 \notag \, .
\eea
where $\dsdeps = \dsdeps \left(\bfeps,\bfepsp,\chi\right) := \partial \bfs / \partial \bfeps$.
Accordingly, $\partial \bfs_{j+1} / \partial \bfeps_{j+1} = \dsdeps \left(\bfeps_{j+1},\bfepsp_{j+1},\chi_{n}\right)$.
Finally, after the FEM update is converged, the plastic strains of the current time step $n+1$ are given as $\bfepsp_{n+1} \leftarrow \bfepsp_{j_{\mathrm{conv}}}$.

\subsection{Update of the density variable}
\label{sec:updateDensity}
Each value of the density field is evaluated for one finite element $e$ as discrete subvolume.
The evolution of the density variable is described by the transient partial differential equation in~\eqref{eq:numericalFieldEquations}$_3$ which needs to be discretized both in time and space for numerical evaluation.
Various strategies can be used for this purpose, \eg{}, a finite element approach would be possible.
However, due to constraint of bounded interval for density $\chi$ and prescribed design volume $\relStructVol$, a direct FE approach consumes a remarkable amount of computation time, cf.~\cite{junker2015variational}, where such a procedure has been discussed.
A more advantageous numerical treatment for this equation has therefore been presented in~\cite{jantos2019accurate} which is based on a generalized FDM along with an operator split.
More details on the numerical performance of this method, also regarding important aspects like convergence behavior and robustness, have been investigated in~\cite{vogel2020adaptive}.
In this paper, we make use of the published finite difference strategies and therefore only recall the fundamental update strategy and refer to the original publications and our published code in \cite{Kick2022Implementation} for a detailed explanation.

The transient character of the evolution equation demands the definition of the initial value for the density variable for each element.
As naive guess, we set each discretized density variable to $\chi_\mathrm{ini} = \relStructVol$.
Therefore, the constraint of the given prescribed structure volume is identically fulfilled.

The change of density is governed by the thermodynamic driving force $\drivingForce$ in equation \eqref{eq:TopOptStrong}.
Considering the operator split, the driving force $\drivingForce$ is based on the Helmholtz free energy \linebreak $\Psi_{0,n+1} := \Psi_{0,n+1} \left( \bfeps_{n+1}, \bfepsp_{n+1}, \chi_{n} \right)$.
High values of the driving force $\drivingForce$ result in increasing densities, and low values result in decreasing densities, respectively.
Since the actual value of the driving force is of no significance, it is thus suitable to normalize the driving force with the weighted driving force (cf. equation (36) in \cite{jantos2019accurate}) by
\be
   \drivingForce_{\mathrm{w}} := \frac{\D \sum_e \left(\chi_{e} - \chi_{\mathrm{min}}\right) \, \left(1 - \chi_{e} \right) \, \drivingForce_{e}}{\D \sum_e \left(\chi_{e} - \chi_{\mathrm{min}}\right) \, \left(1 - \chi_{e} \right)}
\ee
to define the dimensionless driving force $\bar \drivingForce := \drivingForce/\drivingForce_{\mathrm{w}}$.
Equally, we denote the normalized regularization parameter $\bar\beta := \beta / \drivingForce_{\mathrm{w}}$ and the normalized viscosity parameter $\bar\eta := \eta / \drivingForce_{\mathrm{w}}$.

Subsequently, the update scheme is employed according to \cite{jantos2019accurate}.
Then, the discretized evolution equation for the density variable for each element is given by
\be
  \label{eq:updateDensity}
  \chi_{n+1} = \chi_{n} + \Delta t \, \frac{1}{\bar\eta} \, \left[ - \bar \drivingForce_{n+1} + \bar\beta \, \lap\chi_{n} + \lambda_{\mathrm{\chi}} + \gamma \right]
\ee
analogously to equation (49) \cite{Junker2021hyperelastic}.
Here, we are able to account for the regularization parameter~$\bar\beta$ in length unit squared and the viscosity~$\bar\eta$ in time unit as general optimization parameters.

To determine the value of the Lagrange parameter $\lambda_{\mathrm{\chi}}$ for the volume constraint, the update equation \eqref{eq:updateDensity} is solved iteratively by a simple bisection algorithm analogously to Alg.~1 in \cite{jantos2019accurate}.
This process also determines $\gamma$.
Both are implemented in \cite{Kick2022Implementation} with the density update scheme.

\subsection{Optimization process}
\begin{figure}[!htb]
  \centering
  \includegraphics{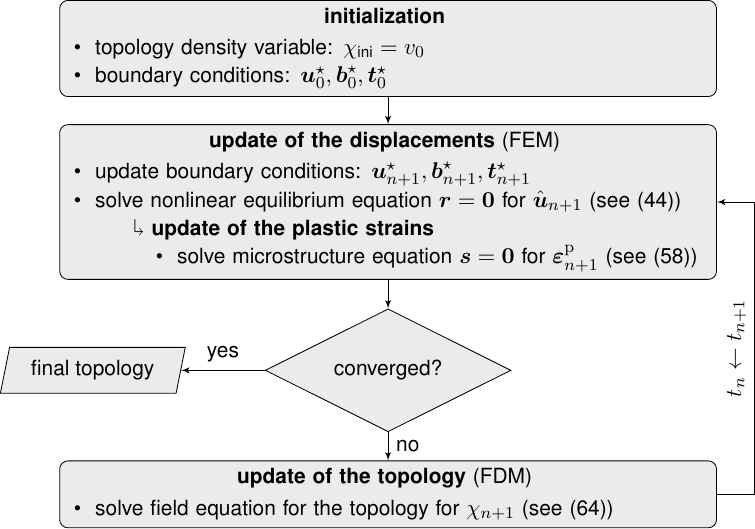}
  \caption{Flowchart of the proposed numerical implementation of the thermodynamic topology optimization including hardening materials.}
  \label{fig:flowchart_TTO}
\end{figure}

The presented update schemes take place in a global optimization process which is given as flowchart in \figref{fig:flowchart_TTO}.
As proposed, we denote this staggered process of FEM and FDM as NEM, cf. \cite{jantos2019accurate}:
first, both the update of the displacements $\bfu_{n+1}$ and the plastic strains $\bfepsp_{n+1}$ are solved by the monolithic finite element method for fixed values of the density variable $\chi_{n}$ at the previous optimization iteration step.
After the FEM has converged, the update of the density variable $\chi_{n+1}$ is performed using the updated displacements $\bfu_{n+1}$ and plastic strains $\bfepsp_{n+1}$.
The updated value for the density variable is used for updating the displacements and plastic strains in the succeeding optimization iteration step $n \leftarrow n+1$.

\FloatBarrier
\section{Numerical results}
\begin{figure}[!htb]
  \centering
  \includegraphics{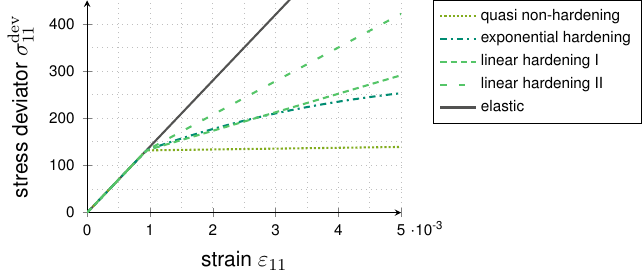}
  \caption{Overview of investigated material behavior: elastic and different types of hardening including models.}
  \label{fig:plot_models_mm_types}
\end{figure}

We present several aspects of our novel thermodynamic topology optimization including hardening materials by investigation of various numerical experiments.
We begin with the presentation of the general functionality of the proposed surrogate material model on the material point level.
Afterwards, we show the impact of the material model on the optimized construction parts by means of analyzing several quasi-2D and 3D boundary value problems.
All results are based on our numerical implementation \cite{Kick2022Implementation} in Julia \cite{Julia}.
We use the material parameter for steel summarized in \tabref{tab:materialParameters}.
\begin{table}[!htb]
  \centering
  \centering
  \caption{Material parameters for steel.}
  \label{tab:materialParameters}
  \begin{tabular}{L{3.1cm}C{0.6cm}R{1.3cm}L{1.0cm}C{0.03mm}}
    Young's modulus & $E$ &  \num{210000} & [\si{\mega\pascal}] \\
    Poisson's ratio & $\nu$ & \num{0.3} & [-] \\
    yield stress & $\sigmaYexp$ & \num{300} & [\si{\mega\pascal}] \\
  \end{tabular}
\end{table}
The yield stress for modelling results from the material parameter with $\sigmaY = \sqrt{2 / 3} \, \sigmaYexp$.
An overview of the different material models used in the following is given in \tabref{tab:materialParametersPlasticity} and \figref{fig:plot_models_mm_types} on material point level.
\begin{table}[!htb]
  \centering
  \caption{Investigated material parameters for hardening including models.}
  \label{tab:materialParametersPlasticity}
  \begin{tabular}{l|ccc}
    hardening models & $h\text{, }h_1 \ [\si{\mega\pascal}]$ & $h_0 \ [\si{\mega\pascal}]$ & $\kappa \ [-]$\\
    \hline \hline
    quasi non-hardening & $0.01\,E$ & -- & -- \\
    exponential hardening & $0.01\,E$ & $0.30\,E$ & \num{300} \\
    linear hardening I & $0.30\,E$ & -- & -- \\
    linear hardening II & $0.80\,E$ & -- & -- \\
  \end{tabular}
\end{table}

\subsection{Surrogate model for hardening materials}
The computation of plastic strains takes place at the microstructural level.
To investigate the results of the proposed surrogate model for hardening materials, we present a first result at the material point and thus without topology optimization.
Consequently, we prescribe the strain as linear function of \num{100} load steps with tension loading and unloading.
For this, we determine the strain tensor depending on the load step $l$ according to
\be
  \label{eq:PrescribedStrain}
  \bfeps(l) = \varepsilon_{11}(l) \, \left(
  \begin{array}{rrr}
    1   & 0.6   & 0.6   \\
    0.6 & -\nu  & -0.1  \\
    0.6 & -0.1  & -\nu  \\
  \end{array} \right) \, .
\ee
To present a result that is representative, the diagonal entries correspond to the material parameters given above (\tabref{tab:materialParameters}), \ie{}, we use the Poisson's ratio of steel, and the shear components have been chosen randomly.
The maximum value of the component in $11$-direction is set to $\varepsilon_{11,\max}(l) = 0.005 \, [-]$.
The numerical results for the surrogate  model at the material point are given as stress/strain diagram exemplary for non-hardening.

\begin{figure}[!htb]
  \centering
 \includegraphics{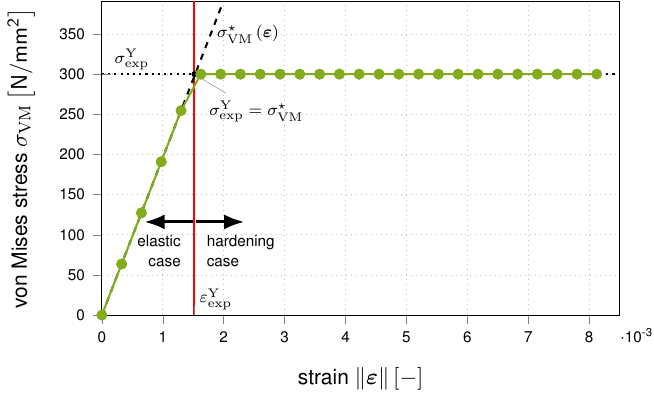}
  \caption{Overview of the decision criteria for microstructural update cases using the surrogate model with non-hardening. No hysteresis occurs due to the motivation of the model, \ie{}, vanishing dissipation function.}
  \label{fig:plot_spmodel_mm_yielding}
\end{figure}

Matching the scalar-valued comparison of the indicator function, the von Mises stresses are plotted above the norm of strains in \figref{fig:plot_spmodel_mm_yielding}.
It indicates that the intended material behavior is displayed: first, the stress/strain curve of the proposed material model increases linearly in the elastic region.
The end points of the elastic region are indicated by $\epsYexp$ and $\sigmaYexp$, respectively.
Then, the stress reaches the yield stress level $r$, here $\sigmaYexp$, in the hardening case.
This behavior coincides to classic plasticity models.
However, the remarkable difference is that the increase or decrease of plastic strains in the surrogate material model directly reacts on the increase or decrease of strains in the hardening case.
To this end, the distinction of elastic and hardening case depends on the strains instead of stresses.
No hysteresis is observed, but with decreasing strains, the stress level is maintained until the strains indicate the elastic region.
The result is thus independent of the unloading history.

\begin{figure}[!htb]
  \centering
  \includegraphics{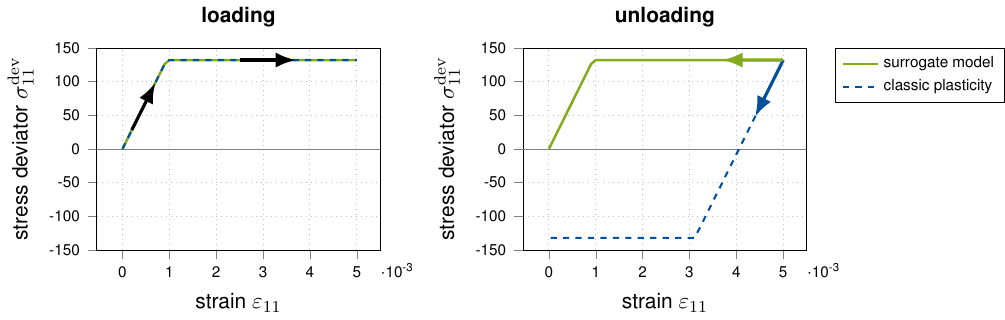}
  \caption{Material point curve of the surrogate model and the classic elasto-plastic model for non-hardening materials and ideal plasticity, respectively. The new idea about the surrogate model is the ``virtual'' unloading with no hysteresis compared to the physical unloading.}
  \label{fig:plot_models_mm_compare}
\end{figure}

In addition to focusing on a special case of plasticity, an important difference of our novel surrogate hardening including model to classic elasto-plastic material models is that we do not formulate our model by using an ordinary differential equation.
Consequently, path-dependence, as intended, is excluded in our model.
Of course, there exists no proof that the different formulations, ODE for classic models vs. algebraic equation for our model, give same results even when only the loading case is considered for which we demand a similar material behavior.
To investigate the quality of our novel surrogate model in this regard, we compare the surrogate material model and the hysteresis curve for a classic elasto-plastic model accounting for one component of the 3D stress/strain state.
Thus, both curves are shown in \figref{fig:plot_models_mm_compare}.
Here, the behavior for loading and unloading can be observed in greater detail.
As a result, the surrogate material model deviates from the purely physical classic elasto-plastic material behavior exactly as intended.
Both models show the identical physical loading path but differ in unloading while the classic model results in the typical hysteresis by dissipation during physical unloading, the virtual unloading follows back the loading path in our surrogate model.
Therefore, the proposed surrogate model displays a reasonable hardening material behavior comparable to physical plasticity but without considering dissipation.

\textit{Remark:} It is worth mentioning that we obtain exactly the behavior as for hyperelastic surrogate model in the 1D case.
However, this holds true for each individual tensor component which differ in different stress levels in the plastic regime which are determined by the specific strain state.
Consequently, our surrogate material model yields the intended results also for the 3D case in which the calibration of a hyperelastic model is a very challenging task, if possible at all.

\begin{figure}[!htb]
  \centering
  \includegraphics{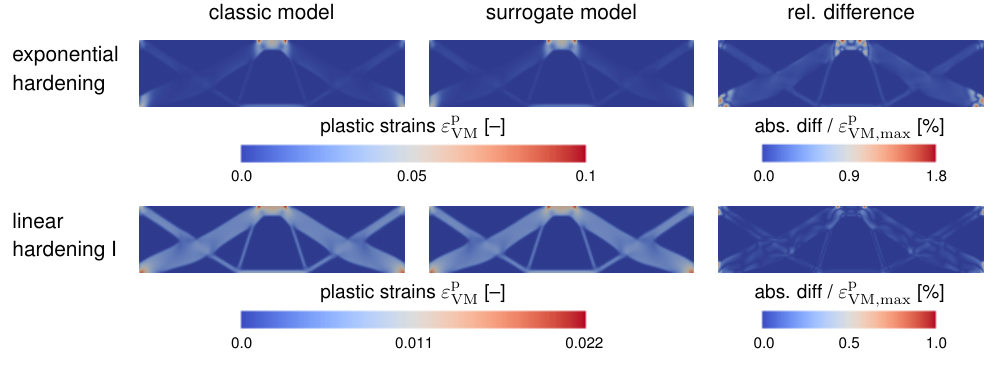}
  \caption{Comparing the classic and the surrogate model by a FEM simulation on a given structure which is loaded in \num{100} steps to the maximum load of $\bfu=\SI{-0.05}{\milli\metre}$ accounting for different hardening characteristics shows very small differences in the plastic strains.}
  \label{fig:pic_models_fem_compare}
\end{figure}

Another investigation of the quality of the surrogate model is discussed by the results of a FEM simulation.
To this end, we choose a fix density distribution of the clamped beam (defined in \secref{sec:benchmarkProblems}) given by the elastic optimization result (\figref{fig:pic_ClampedBeam_tto_compare_chi}).
For this structure and boundary value problem, both for the surrogate model and classic elasto-plasticity a simulation is applied in which we ramp the maximum displacement up over \num{100} load steps.
Computations are performed for exponential and linear hardening I, for instance.
The resulting distribution of plastic strains and its relative difference is plotted in \figref{fig:pic_models_fem_compare}.
The maximum deviation is always less than~\SI{2}{\percent}.
Considering the mathematical difference of the two models, the difference of computed plastic strain is unexpectedly low.

This allows us to validate that the surrogate model along with its implementation address the proposed aspects on the material point level and also confirms accuracy within the FEM.
It is worth mentioning that the surrogate model is proven to represent a special case of classic plasticity for monotonic loading and therefore, without dissipation.

\FloatBarrier
\subsection{Optimization results with surrogate model for hardening materials}
\subsubsection{Benchmark problems and optimization parameters}
\label{sec:benchmarkProblems}
To demonstrate the functionality of the consideration of hardening materials in the thermodynamic topology optimization, several boundary value problems are tested.
To this end, we present all considered design spaces with the respective boundary conditions and symmetry planes.

\begin{figure}[!htb]
  \centering
  \begin{minipage}[b]{0.46\textwidth}
    \centering
    \includegraphics{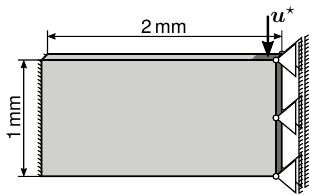}
    \caption{Dimensions of design space and boundary value conditions for the quasi-2D clamped beam.}
    \label{fig:pic_ClampedBeam_bvp}
  \end{minipage}
  \hspace{0.06\textwidth}
  \begin{minipage}[b]{0.46\textwidth}
    \centering
    \includegraphics{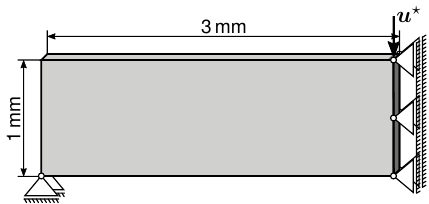}
    \caption{Dimensions of design space and boundary value conditions for the quasi-2D MBB.}
    \label{fig:pic_MBB_bvp}
  \end{minipage}
\end{figure}

\begin{figure}[!htb]
  \centering
  \includegraphics{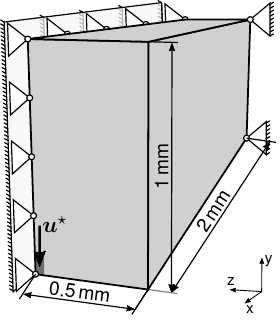}
  \caption{Dimensions of design space and boundary value conditions for the 3D cantilever.}
  \label{fig:pic_3DCantilever_bvp}
\end{figure}

The quasi-2D clamped beam in \figref{fig:pic_ClampedBeam_bvp} is fixated at both sides and chosen in analogy to Maute \etal{}~\cite{Maute1998Adaptive}.
The quasi-2D classical Messerschmitt-Bölkow-Blohm (MBB) beam shown in \figref{fig:pic_MBB_bvp} is simply supported at the lower corner nodes.
Both models are loaded centrally (without symmetry plane) on the design space from above.
As 3D example, we investigate the boundary value problem given in \figref{fig:pic_3DCantilever_bvp} and denote it as 3D cantilever.
The corners of one side are fixated and the load is exerted at the bottom of the opposite side.
All models are discretized by hexahedral finite element meshes with element edge size $e_{\mathrm{size}}$ and linear shape functions.
The thickness of the quasi-2D models is discretized by one finite element with size $e_{\mathrm{size}}$.

It is worth mentioning that in contrast to topology optimization of linear elastic materials, our results depend in a non-linear way on the amplitude of load (which might be provided either by external forces or prescribed displacements).
Here, the load conditions are applied as prescribed displacements $\bfu^{\star}$ where $\bfu^{\star}_{\max}$ is chosen such that plastic strains evolves during optimization.

Our novel surrogate model allows to account for a reasonable computation of the plastic strains without repeating the entire loading history for each optimization step which is usually necessary to estimate the sensitivity.
Therefore, it is worth mentioning that maximum loading, \ie{}, the loading for which the structure is optimized, can be employed instantaneously.
This is a remarkable difference to other optimization schemes including classic plasticity.
Due to this reason, the number of necessary FEM simulations per iteration can be reduced to even one monolithic simulation which significantly reduces the computation time.

The density variable can be varied in the interval $\left[\chi_{\min}, 1\right]$ where the minimum value is set to $\chi_{\min} = \num{0.001}$.
Therefore, the minimal material stiffness is given by $\num{e-9}\times \dsE_0$.
The regularization parameter is chosen as $\bar\beta = 2 \, e_{\mathrm{size}}^2 \, \si{\square\milli\metre}$ and the viscosity for all simulations is set to $\bar\eta = \SI{15}{\second}$, corresponding to our previous work \cite{jantos2019accurate}.
All necessary model and optimization parameters for the different boundary value problems are collected in \tabref{tab:modelParameters}.
\begin{table}[!htb]
  \centering
  \caption{Model and optimization parameters.}
  \label{tab:modelParameters}
  \begin{tabular}{l|rlclccc}
    boundary value problem & \#$_\mathrm{elements}$ & $e_{\mathrm{size}} \ [\si{\milli\metre}]$ & $\bfu^{\star} \ [\si{\milli\metre}]$  & $\relStructVol \ [-]$ & $\bar\eta \ [\si{\second}]$  & $\bar\beta \ [\si{\square\milli\metre}]$ \\
    \hline \hline
    quasi-2D clamped beam & \num{5000}\hspace*{2.0mm} & \hspace*{4.8mm}\num{0.020} & \num{0.05} & \hspace*{1.8mm}\num{0.41} & \num{15} & \num{0.80e-3} \\
    quasi-2D MBB beam & \num{4800}\hspace*{2.0mm} & \hspace*{4.8mm}\num{0.025} & \num{0.02} & \hspace*{1.8mm}\num{0.5} & \num{15} & \num{1.25e-3} \\
    3D cantilever & \num{26364}\hspace*{2.0mm} & \hspace*{4.8mm}\num{0.038} & \num{0.06} & \hspace*{1.8mm}\num{0.14} & \num{15} & \num{2.96e-3} \\
  \end{tabular}
\end{table}

The illustrations of the field data are created with Paraview \cite{Paraview}.
Even if the models make use of symmetry planes, the results are presented as whole (mirrored) in some instances.
The resultant structures are obtained by using the isovolume filter for the density variable $\chi$ with the minimum threshold set to \num{0.5}.
This is the average value of the interval in which $\chi$ has been defined.

\FloatBarrier
\subsubsection{Optimal structures}
\label{sec:optimizedStructures}
We investigate the impact of inclusion of the behavior of hardening materials on the resultant optimal structure.
To this end, the optimization results are compared with results of thermodynamic topology optimization for a linear elastic material behavior.
This can be achieved while setting the yield stress to an unphysically high value, \ie{}, $\sigmaYexp = \SI{500000}{\mega\pascal}$.
This ensures that no plastic deformation is active since the von Mises norm of the stress is below this value for all boundary value problems considered.
The results obtained from this elastic optimization are, of course, consistent with results obtained in our previous publications, cf. \cite{jantos2019accurate}, for instance.
All structures are presented for the converged iteration step $n_{\mathrm{conv}}$.
The structures with shades of green correspond to the thermodynamic topology optimization including hardening materials whereas the gray structure is the result for a purely linear elastic topology optimization.
For examples of the side-by-side evolution of topology and plastic strains, we refer to \secref{sec:evolutionProcess}.

\begin{figure}[!htb]
  \centering
    \includegraphics{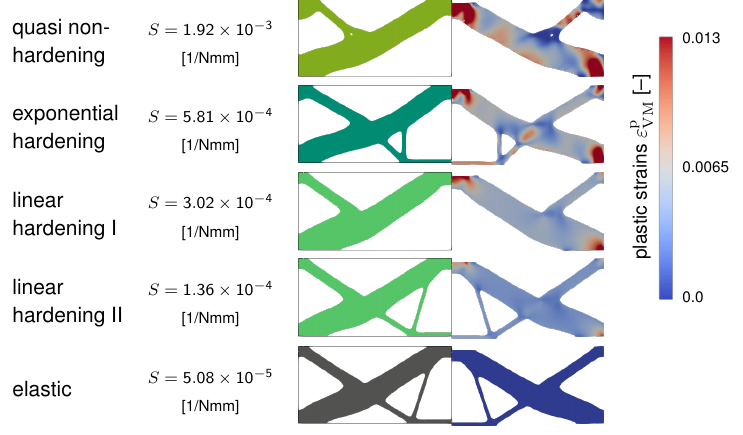}
  \caption{Overview of optimized structures with stiffness, plastic strains and deformation for the clamped beam with the surrogate material model accounting for different hardening characteristics and linear elasticity.}
  \label{fig:pic_ClampedBeam_tto_compare_chi}
\end{figure}

\begin{figure}[!htb]
  \centering
  \includegraphics{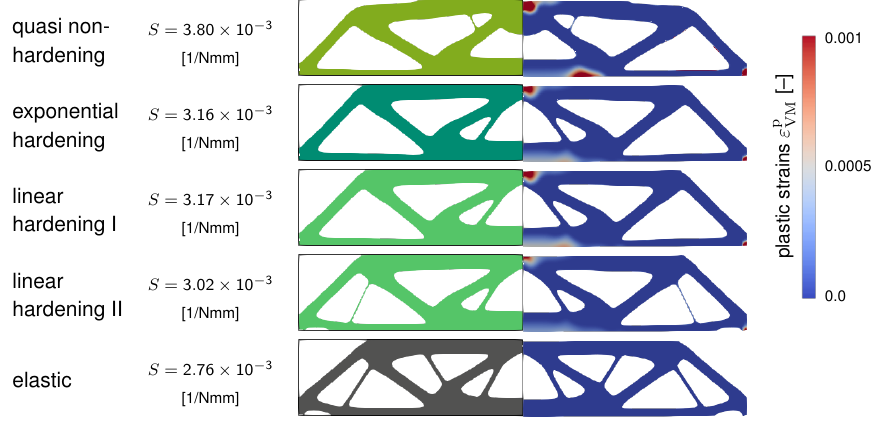}
  \caption{Overview of optimized structures with stiffness, plastic strains and deformation for the MBB with the surrogate material model accounting for different hardening characteristics and linear elasticity.}
  \label{fig:pic_MBB_tto_compare_chi}
\end{figure}

Due to loading, high plastic strains may occur in the entire design space.
Each area with plastic deformations results in the greatest stresses limited by the yield threshold $r$.
In turn, the rearrangement of densities allows lower stress intensities in topology: i) thicker cross-section areas reduce the maximum value of the averaged stress such that the remaining stress is limited by the yield criterion $r$, or ii) vanishing substructures.
Here, the analysis of the optimal structures reveals exactly the mentioned effects in the resulting topologies.
One the one hand, we observe thicker member sizes, cf. the supports of the clamped beam in \figref{fig:pic_ClampedBeam_tto_compare_chi} and the center of the quasi non-hardening MBB in \figref{fig:pic_MBB_tto_compare_chi}, for instance.
On the other hand, fewer trusses can be detected for results optimized with hardening materials in general.
Here, also the distribution of plastic strains are shown for each final structure.
Further, it is noticeable that high plastic strains are clipped in order for a good representation of the distribution of plastic strains for each hardening type.
This final distribution of plastic strains shows that the highest stresses and thus plastic deformations are present at the constrained boundaries in terms of external loading and supports.
Moreover, high plastic strains occur in the complete middle part of the design space of the clamped beam , cf. \secref{sec:evolutionProcess}.
For this reason, a large void area below the truss corresponding to the loading can be seen in \figref{fig:pic_ClampedBeam_tto_compare_chi}.
This void region is even wider than in the elastic optimization result.
This has an impact on the total structure regarding the general topology: due to prescribed total structure volume, the angles and thicknesses of some bars change significantly for optimization with hardening including models.
Consequently, remarkably different structures are computed when hardening material behavior is considered.

The special feature to define different characteristics of hardening with our approach allows us to evaluate its impact on optimal structures.
In general, differences are obvious between optimal structures including hardening materials, cf. for the clamped beam in \figref{fig:pic_ClampedBeam_tto_compare_chi} and for the MBB in \figref{fig:pic_MBB_tto_compare_chi}.
Especially, the quasi non-hardening structure stands out strongly from those with significantly more hardening.
Particularly interesting is the impact observed in the results with an increasing hardening slope.
The higher the magnitude of hardening, the closer the material behavior belongs to linear elastic behavior.
This can be observed, \eg{}, for the MBB in \figref{fig:pic_MBB_tto_compare_chi}, in the vertical order.
Obviously, the structure with linear hardening II approaches the elastic result: additional trusses and the notch near the supports are created.
This structural changes correspond to the noted stiffness $S$ values which are investigated in \secref{sec:ConvergenceBehavior} in more detail.
Comparing the results with exponential hardening and linear hardening I, interestingly, the structures of the clamped beam are clearly different according to the different types of hardening whereas both structures of the MBB are almost the same.
The difference is due to the fact that the prescribed loading for the clamped beam leading to higher strains and therefore larger differences for the yield criterion.
The further the yield criterion differs from the start point of yielding, the more the characteristic of hardening influences the topology and greater structural deviations become apparent.
In general, these results suggest that a precisely defined material behavior affect the optimal structure.

\begin{figure}[!htb]
  \centering
  \includegraphics{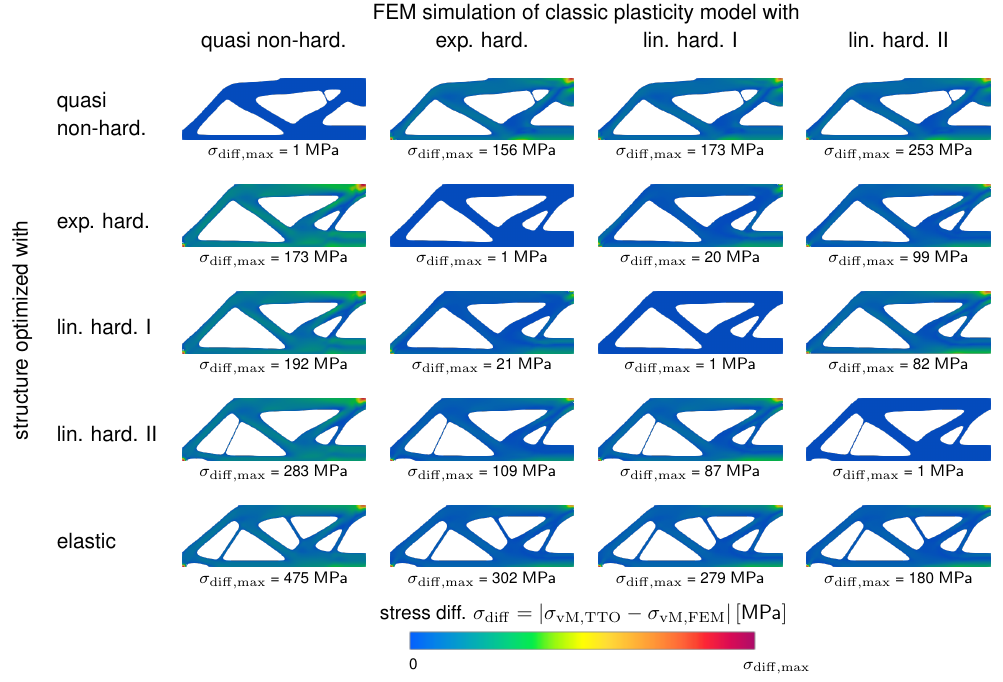}
  \caption{Cross-check influence of correct material model due to FEM simulations where all optimized MBB structures are loaded in \num{100} steps to the maximum of $\bfu=\SI{-0.02}{\milli\metre}$ with respect to different types of hardening.}
  \label{fig:pic_MBB_model-cross-check_stress}
\end{figure}

The influence of including an incorrect material model during optimization is cross-checked by further simulations.
For this purpose, the structures optimized with any surrogate hardening material model are simulated by FEM regarding classic plasticity of all types of hardening considered here, cf. \figref{fig:pic_MBB_model-cross-check_stress}.
Remarkably, a difference of less than \SI{1}{\mega\pascal} occurs for simulations when the related surrogate model for each hardening type is used already during optimization.
This great agreement confirms once again the comparability of our surrogate model including hardening as a special case of classic elasto-plastic models.
Bigger stress differences are detected for a larger difference in hardening type.
Especially, results with quasi non-hardening show large stress deviations from the simulation results with other types of hardening or elasticity.
Since the curves for the exponential and the linear hardening I regarding smaller loads are very similar, the deviations in these cross-check simulations are also small.
Regions of the structure where large differences in the stresses between different material models are visible do not represent an optimal structure considering the real material behavior.
It is observed that large differences often occur at loading areas and support. 
Here, large internal energies with the associated risk for failure are present.
Evidently, the inclusion of the real material behavior in the topology optimization is important.

\begin{figure}[!htb]
  \centering
  \includegraphics{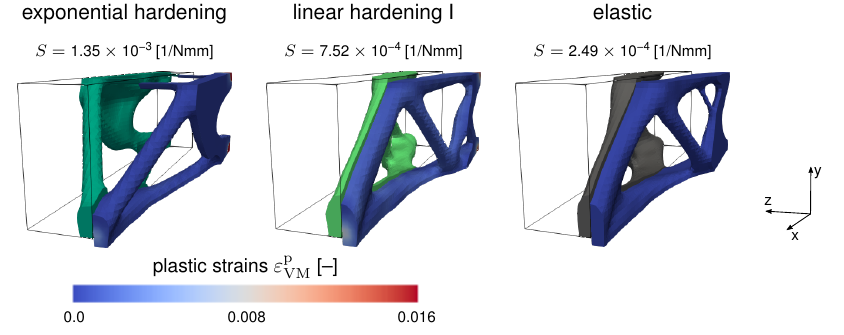}
  \caption{Examples of optimized structures with stiffness, plastic strains and deformation for the 3D cantilever with the surrogate material model accounting for different hardening characteristics and linear elasticity.}
  \label{fig:pic_3DCantilever_tto_compare_chi}
\end{figure}

Further optimizations prove the functionality and applicability of our approach for fully 3D boundary problems: the 3D cantilever is optimized, for example, with exponential hardening and linear hardening I.
The resulting optimal 3D structures are shown in \figref{fig:pic_3DCantilever_tto_compare_chi}.
Similar to the quasi-2D structures, the differences between elastic and different hardening characteristics are obvious.

\FloatBarrier
\subsubsection{Convergence behavior}
\label{sec:ConvergenceBehavior}
Another aspect of analysis is to discuss the evolution of the optimization objective which is to minimize the compliance of the structure.
Since a compliance minimization analogously causes a stiffness maximization, we use the latter for presentation.
The stiffness is computed in analogy to other works on topology optimization by $S = 1/ (\bff\cdot\hat{\bfu})$.
Consequently, we expect a decreasing function for $S$ when the reaction force increases during the evolution of the structure.
The order of magnitude of stiffness is very different for elastic and hardening materials optimization.
For a convincing representation a logarithmic stiffness axis is chosen.
We define convergence as soon as the relative stiffness changes less than \num{1e-5} for the first time and less than \num{1e-4} for further two succeeding iteration steps.
This rather strict convergence criterion is chosen to exclude a wrong detection of convergence for optimizations with hardening materials.

\begin{figure}[!htb]
  \centering
  \includegraphics{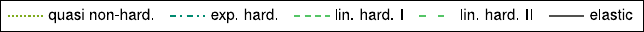}\\
  \vspace*{2mm}
  \begin{subfigure}[b]{0.49\textwidth}
    \includegraphics{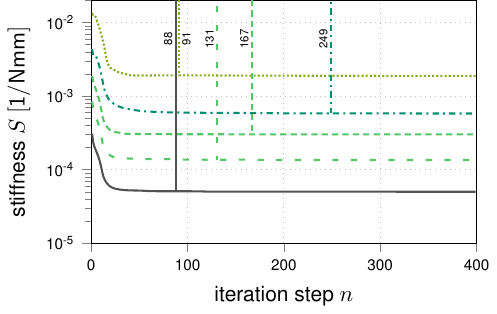}
    \caption{clamped beam}
    \label{fig:plot_ClampedBeam_u3_stiffness}
  \end{subfigure}
  \begin{subfigure}[b]{0.49\textwidth}
    \includegraphics{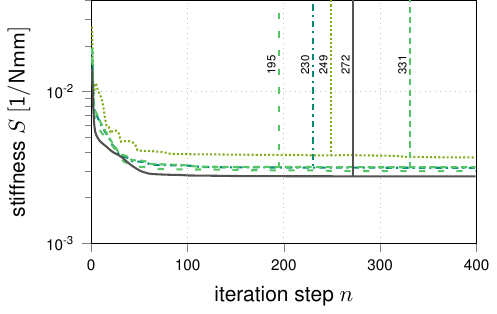}
    \caption{MBB}
    \label{fig:plot_MBB_u3_stiffness}
  \end{subfigure}
  \caption{Convergence of stiffness evolution during the optimization process.
  The first time that the convergence criterion has been reached is indicated by a vertical line.}
  \label{fig:plot_u3_stiffness}
\end{figure}

The stiffness and iteration step of convergence is plotted for the clamped beam and the MBB in \figref{fig:plot_u3_stiffness}.
We still see the usual evolution of the stiffness during topology optimization which is that the stiffness increases while a discrete black/white structure evolves.
In the end, the maximum stiffness converges towards a constant value.

The onset of hardening includes remarkable reduction of stiffness since locally higher strains do not result in higher stress: the yield stress $\sigmaY$ is the fixed upper limit for non-hardening, and the increase of stress is slowed down with hardening which is a reasonable behavior.
Therefore, the stiffness of structures accounting for the behavior of hardening materials is lower than of those which behave purely elastically.
This becomes particularly clear with the clamped beam in \figref{fig:plot_ClampedBeam_u3_stiffness} where larger values of plastic strains are observed.
In general, the (absolute value of the) differences in the stiffness plots corresponds to the dissipated energy due to the plastic formation of deformations.

Furthermore, the plastic strains are even lower for hardening than for quasi non-hardening.
This is caused by the yield criterion $r$ which allows the stresses to increase in a defined manner with hardening.
Therefore, the plots also show a greater stiffness during the increase of the hardening parameters with exponential, linear hardening I and II.
Structures with a higher stiffness are thus more similar to elastically optimized structures, cf. the clamped beam in \figref{fig:pic_3DCantilever_tto_compare_chi} with linear hardening II.

It is remarkable that sometimes optimizations with hardening materials converge in less iteration steps than the elastic optimizations, cf. \figref{fig:plot_u3_stiffness} especially for the MBB.
The number of convergence iterations is a major factor for the difference in computation time in hardening including and elastic optimizations.
Additionally, our novel surrogate model enables us to save numerical cost by reducing the number of necessary FEM simulations per iteration due to the missing path-dependence.
Nevertheless, it is obvious that computation time also increases with the number of Newton iterations within the monolithic FEM simulations.
Both can be seen by comparing the runtimes for elastic and hardening material including optimizations in \tabref{tab:convergenceRuntime}.
\begin{table}[!htb]
  \centering
  \caption{Convergence iteration steps, average of the necessary FEM iterations per optimization step and relative runtime according to elastic optimization.}
  \label{tab:convergenceRuntime}
  \begin{tabular}{l|lrrr}
    boundary value problem & type of plasticity & \makecell{conv. iteration\\step $n_{\mathrm{conv}}$} & \makecell{avg. FEM\\iterations $\bar j$} & rel. runtime \\
    \hline \hline
     & elastic & \num{88}\hspace*{8.5mm} & \num{1.00}\hspace*{6.0mm} & \num{1.00}\hspace*{6.0mm} \\
     & quasi non-hardening & \num{91}\hspace*{8.5mm} & \num{30.64}\hspace*{6.0mm} & \num{23.02}\hspace*{6.0mm}  \\
     & exp. hardening & \num{249}\hspace*{8.5mm} & \num{4.14}\hspace*{6.0mm} & \num{13.36}\hspace*{6.0mm}  \\
     & linear hardening I & \num{167}\hspace*{8.5mm} & \num{2.59}\hspace*{6.0mm} & \num{5.90}\hspace*{6.0mm}  \\
    \multirow{-5}{*}{quasi-2D clamped beam} & linear hardening II & \num{131}\hspace*{8.5mm} & \num{2.39}\hspace*{6.0mm} & \num{4.55}\hspace*{6.0mm}  \\
    \hline
     & elastic & \num{272}\hspace*{8.5mm} & \num{1.00}\hspace*{6.0mm} & \num{1.00}\hspace*{6.0mm} \\
     & quasi non-hardening & \num{249}\hspace*{8.5mm} & \num{4.41}\hspace*{6.0mm} & \num{3.02}\hspace*{6.0mm}  \\
     & exp. hardening & \num{230}\hspace*{8.5mm} & \num{3.24}\hspace*{6.0mm} & \num{2.02}\hspace*{6.0mm}  \\
     & linear hardening I & \num{331}\hspace*{8.5mm} & \num{2.95}\hspace*{6.0mm} & \num{3.10}\hspace*{6.0mm}  \\
    \multirow{-5}{*}{quasi-2D MBB} & linear hardening II & \num{195}\hspace*{8.5mm} & \num{2.92}\hspace*{6.0mm} & \num{1.87}\hspace*{6.0mm}  \\
  \end{tabular}
\end{table}
Here, the average of FEM iterations for optimizations accounting for hardening including models is typically less than \num{5} and therefore significantly lower than the cost for additional FEM simulations for incremental loading in each optimization iteration with classic plasticity models.
Therefore, with the surrogate model the needed computational resources for a optimization including hardening materials is comparable to an elastic optimization in general which is applicable in engineering practice.
The runtime and number of average FEM iterations are noticeably higher for the clamped beam with quasi non-hardening material.
The reason for this exception is the need for additional damping activations.

\FloatBarrier
\subsubsection{Structure evolution during the optimization process}
\label{sec:evolutionProcess}
\begin{figure}[!htb]
  \centering
  \includegraphics{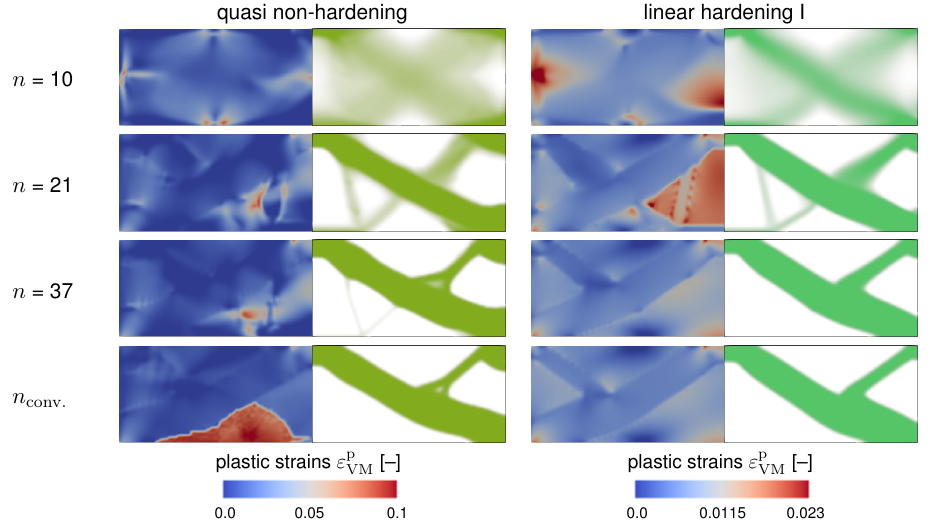}
  \caption{Evolution of increasing/decreasing plastic strains and structure during the optimization process for the clamped beam with quasi non-hardening and linear hardening I.}
  \label{fig:pic_ClampedBeam_tto_evolution}
\end{figure}

\begin{figure}[!htb]
  \centering
  \includegraphics{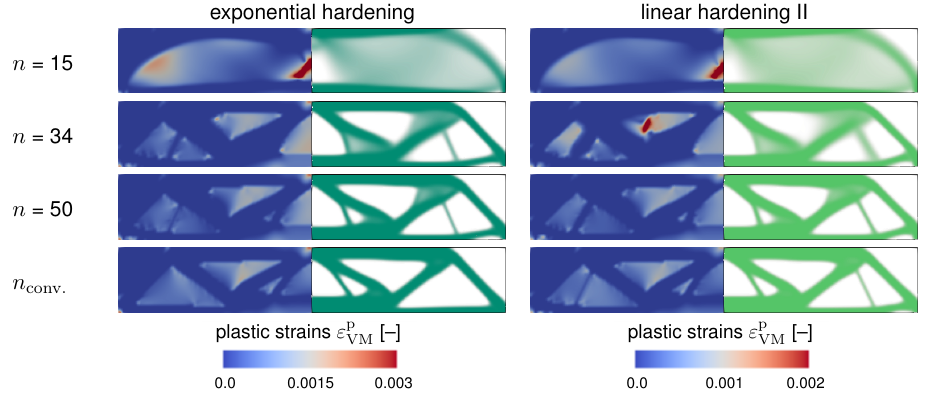}
  \caption{Evolution of increasing/decreasing plastic strains and structure during the optimization process for the MBB with exponential and linear hardening II.}
  \label{fig:pic_MBB_tto_evolution}
\end{figure}

The evolution of the structure and the plastic strains $\bfepsp$ during the optimization process is exemplary presented for the clamped beam with quasi non-hardening and linear hardening I in \figref{fig:pic_ClampedBeam_tto_evolution}
and for the MBB with exponential and linear hardening II in \figref{fig:pic_MBB_tto_evolution}.
For all examples, the structure and void areas under the influence of plastic strains can be observed, as explained in \secref{sec:optimizedStructures}.
Furthermore, we see that the value of plastic von Mises strains on the clamped beam is lower for linear hardening I than for quasi non-hardening which corresponds with the different yield criterions $r$ that are active for defined hardening material characteristics.

It is worth mentioning that the amount of plastic strains also reduces during the optimization while stiffness increases and thus strains are locally reduced (again).
This can be seen when comparing the iteration steps $n=21$ and $n=37$ at the area of external displacement and support for the clamped beam in \figref{fig:pic_ClampedBeam_tto_evolution} as well as for step $n=15$, $n=34$ and $n=50$ at the middle of increasing void areas for the MBB in \figref{fig:pic_MBB_tto_evolution}.
Therefore, it is a crucial property of the proposed material model to reduce plastic strains without dissipation during virtual unloading.
This proves that the proposed surrogate material model for hardening material behavior operates as expected during the optimization process.
It is thus possible to consider the plastic strain evolution simply by considering the current strain while avoiding the repeated computation of the entire loading path.

\section{Conclusions}
A novel approach to the thermodynamic topology optimization including hardening materials was presented.
In order to be able to consider materials that harden after exceeding the initial yield criterion in topology optimization, a novel surrogate material model was developed.
To this end, the model was constructed as special case of classic elasto-plastic models which are dissipation-free such that the plastic strains result from pure energy minimization and path-dependence is lost in the surrogate model.
The resultant system of governing equations followed as stationarity conditions from an extended Hamilton functional.
The system comprised the field equations for the displacements and the density variable, and an algebraic equation for the plastic strains.
In the algebraic equation with respect to the plastic strain tensor, arbitrary types of hardening can be included by defining the related yield criterion: exemplary we used non-hardening, linear hardening and exponential hardening.
For the numerical implementation, we employed the neighbored element method for solving the weak form of the balance of linear momentum and the strong form of the evolution equation for the density variable.
Thereby, optimization is solved as evolutionary process in a staggered manner.
We presented both a material point and FEM investigation to demonstrate the general functionality of the novel material model and various finite boundary value problems for optimization.
Significant deviations between optimized structures for purely elastic materials and the surrogate model for plastic deformations could be detected.
Also differences can be observed with quasi non-hardening, linear hardening or exponential hardening.
All optimizations result in reliable convergence and with a small number of iteration steps.
During the optimization process, our surrogate material model allows both to predict the microstructural state both for increasing and decreasing strain states due to topology optimization: the plastic strains always correspond to a state evolved during pure loading as is the case for the optimized component during real application.
Thus, the surrogate model for hardening materials in thermodynamic topology optimization provides reasonable optimization results for real-world material parameters which are obtained with low computational effort.

\FloatBarrier
\section*{Acknowledgment}
We highly acknowledge the funding of this research by the Deutsche Forschungsgemeinschaft (DFG, German Research Foundation) through the project grant JU 3096/2-1.

\section*{Conflict of interest}
On behalf of all authors, the corresponding author states that there is no conflict of interest.

\section*{Ethical approval}
Not applicable.

\section*{Replication of results}
The results can be replicated following the algorithms provided in the paper. Furthermore, the implementation in Julia is available as indicated in~\cite{Kick2022Implementation}.

\appendix
\section{Derivation of the surrogate material model}
\label{appendix:DerivationSurrogateModel}
From the stationarity condition \eqref{eq:generalFieldEquations}$_2$, the Lagrange parameters $\lambda_\mathrm{\sigma}$ and $\lambda_\mathrm{V}$ need to be computed.
Therefore, let us reformulate \eqref{eq:StatPlastLoc} such that we can compute $\lambda_\mathrm{\sigma}$ and $\lambda_\mathrm{V}$ analytically.
To this end, both sides of \eqref{eq:StatPlastLoc} are double contracted by the deviator operator $\dsP$ from the left hand side.
This yields
\be
  - \bfsigmadev - \lambda_\mathrm{\sigma} \left[ \frac{\bfsigmadev}{\norm{\bfsigmadev}} :\dsP: \dsEChi + \dsP : \pf{r}{\bfepsp} \right] = 0
\ee
where we used $\dsP:\bfsigmadev=\bfsigmadev$ and $\dsP:\bfI = \boldsymbol{0}$.
Furthermore, it holds $\bfsigmadev:\dsP: \dsEChi = \bfsigmadev: \dsEChi$.
Afterwards, we double contract both sides by the stress deviator $\bfsigmadev$ from the right-hand side, yielding
\be
  - \bfsigmadev:\bfsigmadev - \lambda_\mathrm{\sigma} \left[ \frac{\bfsigmadev}{\norm{\bfsigmadev}} : \dsEChi : \bfsigmadev + \dsP : \pf{r}{\bfepsp} : \bfsigmadev \right] = 0 \, .
\ee
Finally, we insert the constraint $\bfsigmadev:\bfsigmadev = \norm{\bfsigmadev}^2 = r^2$ and $\norm{\bfsigmadev}=r$, respectively, and also account for $\dsP:\bfsigmadev=\bfsigmadev$ which gives us
\be
  \lambda_\mathrm{\sigma} = - \frac{r^{3}}{\D \bfsigmadev : \dsEChi : \bfsigmadev + \pf{r}{\bfepsp} : \bfsigmadev \, r} \, .
\ee
To compute the Lagrange parameter $\lambda_\mathrm{V}$, we double contract \eqref{eq:StatPlastLoc} with $\bfI$ from the right-hand side.
This results in
\be
  - \bfsigma : \bfI - \lambda_\mathrm{V} \, \bfI : \dsEChi : \bfI = 0
\ee
where we used $\dsP : \bfI = \boldsymbol{0}$ and $\partial r / \partial \bfepsp = \boldsymbol{0}$ in case of non-hardening or
\be
  \pf{r}{\bfepsp} : \bfI = \pf{\rH}{\norm{\bfepsp}} \, \pf{\norm{\bfepsp}}{\bfepsp} : \bfI = \pf{\rH}{\norm{\bfepsp}} \, \frac{\bfepsp : \bfI}{\norm{\bfepsp}} = r' \, \frac{\bfepsp : \bfI}{\norm{\bfepsp}} = 0
\ee due to $\bfI : \bfepsp = 0$, respectively.
Inserting the constraint $\bfsigma : \bfI = \bfI : \dsEChi : \bfeps$ results in
\be
  \lambda_\mathrm{V} = - \frac{\bfI : \dsEChi : \bfeps}{\bfI : \dsEChi : \bfI} \, .
\ee
Then, we finally find
\bea
  \bfs &:=& - \bfsigma + \frac{r^3}{\D \bfsigmadev :  \dsEChi : \bfsigmadev + \pf{r}{\bfepsp} : \bfsigmadev \, r} \, \left[\frac{\bfsigmadev:\dsEChi}{r} + \pf{r}{\bfepsp} \right] \\
  && + \frac{\bfI : \dsEChi : \bfeps}{\bfI : \dsE_0 : \bfI} \, \bfI : \dsE_0 = \boldsymbol{0} \notag
\eea
which constitutes as the governing equation for the plastic strains, cf. \eqref{eq:s_epsP}.

\section{Finite element method according to Ferrite.jl}
\label{appendix:FerriteFEM}
A possible implementation of the thermodynamic topology optimization including hardening materials by use of the Ferrite.jl package \cite{Ferrite} and the Tensors.jl \cite{JuliaTensors} is presented in the \algoref{alg:FerriteFEM1} and \algoref{alg:FerriteFEM2}.
This algorithm is deduced from our published code in \cite{Kick2022Implementation}.
\begin{singlespace}
\begin{algorithm}[H]
  \caption{Finite element implementation in Ferrite.jl \cite{Ferrite}}
  \input{./03_figures/alg_Ferrite_fem.tex}
  \label{alg:FerriteFEM1}
\end{algorithm}
\begin{algorithm}[H]
  \caption{Continuation of the finite element method in Ferrite.jl \cite{Ferrite}}
  \input{./03_figures/alg_Ferrite_fem_2.tex}
  \label{alg:FerriteFEM2}  
\end{algorithm}
\end{singlespace}

\addcontentsline{toc}{section}{References}
\bibliographystyle{plain}
\bibliography{bib_short}

\end{document}

%% file: 03_figures/alg_update_epsP.tex
\begin{algorithmic}[]
    \State \State initialize $i=0$ \Comment{initial Newton iterator}
    \State \State initialize $\bfepsp_{\mathrm{ini}} = \boldsymbol{0}$  \Comment{initial plastic strains}
    \State \While{true}
    \State \If{$i > i_{\max}$} \Comment{check maximal loops}
    \State break
    \EndIf
    \State \State compute $\D \tilde\bfs_{i} (\bfepsp_{i})$, see \eqref{eq:s_epsP_scaled} \Comment{residual vector}
    \State \If{$\D \tilde\bfs_{i} (\bfepsp_{i}) = \boldface{0}$} \Comment{check convergence}
    \State break
    \EndIf
    \State \State compute $\D \pf{\tilde\bfs_{i}}{\bfepsp_{i}}$, see \eqref{eq:ds_epsP_scaled} \Comment{analytical derivative}
    \State \State update $\D \bfepsp_{j+1} \leftarrow \bfepsp_{i+1} = \bfepsp_{i} - \left[ \pf{\tilde\bfs_{i}}{\bfepsp_{i}} \right]^{-1} \cdot \tilde\bfs_{i}$, see \eqref{eq:NewtonUpdateESepsp} \Comment{new plastic strains}
    \State \State update $i = i + 1$ \Comment{next Newton iterator}
    \State \EndWhile
\end{algorithmic}

%% file: 03_figures/alg_Ferrite_fem.tex
\begin{algorithmic}[]
    \State \While{true}
      \State \ForEach{$element \in mesh$} \Comment{repeat for each element}
        \State \State call $\text{reinit!}(mesh,elementvalues)$ \Comment{element values}
        \State \ForEach{$ip$ $\in$ $element$} \Comment{repeat for each integration point}
        \State \State compute $\bfeps = \text{function\_symmetric\_gradient}(elementvalues,ip,\bfu_{e})$  \Comment{strains}
        \State \State compute $\bfsigma$, $\bfepsp$ and $\dsD_0$ \Comment{material state and tangent operator}
        \State \State compute $\Omega^{\star} = \text{getdetJdV}(elementvalues, ip)$ \Comment{weighted volume fraction}
        \State \For{$i$ to number base shape functions} \Comment{repeat for number of \begin{flushright} base shape functions \end{flushright}}
          \State \State compute $\bfB^{\mathrm{T}} = \text{shape\_symmetric\_gradient}(elementvalues, ip, i)$ \State \Comment{derivative of shape functions}
          \State \State compute $\bfr_{e}\left[i\right] += \left(\bfB^{\mathrm{T}} : \bfsigma \right) \, \Omega^{\star}$ \Comment{element residual vector}
          \State \For{$j$ to number base shape functions} \Comment{repeat for number of \begin{flushright} base shape functions \end{flushright}}
          \State \State compute $\bfB = \text{shape\_symmetric\_gradient}(elementvalues, ip, j)$  \State \Comment{derivative of base shape function}
          \State \State compute $\bfK_{e}\left[i,j \right] += \left(\bfB^{\mathrm{T}} : \dsD_0 : \bfB \right) \ \Omega^{\star}$ \Comment{element stiffness matrix}
        \State \EndFor
      \State \EndFor
      \State \State compute $\bfr_e = \bfr_e - \bff_{e,\mathrm{ext}}$ \Comment{apply external forces}
    \State \EndFor
    \State \State call $\text{assemble!}(assembler,\bfK_{e},\bfr_{e})$ \Comment{global stiffness matrix, global residual vector}
    \State \EndFor
    \algstore{part1}
\end{algorithmic}

%% file: 03_figures/alg_Ferrite_fem_2.tex
\begin{algorithmic}[]
    \algrestore{part1}
      \State \State call $\text{apply\_zero!}(\bfK,\bfr,constraints)$ \Comment{apply boundary conditions}
      \State \If{$\norm{\bfr} < \text{tol}$} break \EndIf \Comment{check convergence criterion}
      \State \State update $\D \bfu_{i+1} = \bfu_{i} - \frac{\bfr}{\bfK}$ \Comment{displacement vector}
      \State \State update $i = i + 1$ \Comment{next Newton iterator}
    \State \EndWhile
\end{algorithmic}

%% file: TTO_hardening.bbl
\begin{thebibliography}{10}

\bibitem{Alberdi2017Topology}
Ryan Alberdi and Kapil Khandelwal.
\newblock Topology optimization of pressure dependent elastoplastic energy
  absorbing structures with material damage constraints.
\newblock {\em Finite Elements in Analysis and Design}, 133:42--61, 2017.

\bibitem{Amir2016StressConstrained}
Oded Amir.
\newblock Stress-constrained continuum topology optimization: a new approach
  based on elasto-plasticity.
\newblock {\em Struct Multidisc Optim}, 55:1797--1818, 2016.

\bibitem{Paraview}
Utkarsh Ayachit.
\newblock {\em {The ParaView Guide: A Parallel Visualization Application},
  www.paraview.org}.
\newblock Kitware, 2015.

\bibitem{bartels2021cahn}
Alexander Bartels, Patrick Kurzeja, and J{\"o}rn Mosler.
\newblock Cahn--hilliard phase field theory coupled to mechanics: Fundamentals,
  numerical implementation and application to topology optimization.
\newblock {\em Computer Methods in Applied Mechanics and Engineering},
  383:113918, 2021.

\bibitem{Bendsoe1989Optimal}
M.~P. Bends{\o}e.
\newblock Optimal shape design as a material distribution problem.
\newblock {\em Structural Optimization}, 1:193--202, 1989.

\bibitem{bendsoe2003Topology}
M.~P. Bends{\o}e and O.~Sigmund.
\newblock {\em Topology Optimization: Theory, Methods and Applications.}
\newblock Springer-Verlag Berlin Heidelberg, 2003.

\bibitem{Julia}
Jeff Bezanson, Alan Edelman, Stefan Karpinski, and Viral~B Shah.
\newblock {Julia: A fresh approach to numerical computing}, www.julialang.org.
\newblock {\em SIAM {R}eview}, 59(1):65--98, 2017.

\bibitem{Bogomolny2012Conceptual}
Michael Bogomolny and Oded Amir.
\newblock Conceptual design of reinforced concrete structures using topology
  optimization with elastoplastic material modeling.
\newblock {\em International Journal for Numerical Methods in Engineering},
  90(13):1578--1597, 2012.

\bibitem{Bruggi2012Topology}
Matteo Bruggi and Pierre Duysinx.
\newblock Topology optimization for minimum weight with compliance and stress
  constraints.
\newblock {\em Struct Multidisc Optim}, 46:369--384, 2012.

\bibitem{Ferrite}
Kristoffer Carlsson, Fredrik Ekre, and Contributors.
\newblock {Ferrite.jl (Julia package)}, version: 0.3.0, date-released:
  2021-03-25, https://github.com/ferrite-fem/ferrite.jl.

\bibitem{JuliaTensors}
Kristoffer Carlsson, Fredrik Ekre, and Contributors.
\newblock {Tensors.jl (Julia package)}, version: 1.6.1, date-released:
  2021-09-07, https://github.com/ferrite-fem/tensors.jl.

\bibitem{Deaton2014ASurvey}
Joshua~D. Deaton and Ramama~V. Grandhi.
\newblock A survey of structural and multidisciplinary continuum topology
  optimization: post 2000.
\newblock {\em Structural and Multidisciplinary Optimization}, 49:1--38, 2014.

\bibitem{Duester2001ThePVersion}
A.~Düster and E.~Rank.
\newblock The p-version of the finite element method compared to an adaptive
  h-version for the deformation theory of plasticity.
\newblock {\em Computer Methods in Applied Mechanics and Engineering},
  190(15-17):1925--1935, 2001.

\bibitem{Duysinx1998Topology}
P.~Duysinx and M.~P. Bends{\o}e.
\newblock Topology optimization of continuum structures with local stress
  constraints.
\newblock {\em International Journal for Numerical Methods in Engineering},
  43:1453--1478, 1999.

\bibitem{Duysinx1999Topology}
Pierre Duysinx.
\newblock Topology optimization with different stress limits in tension and
  compression.
\newblock {\em Third World Congress of Structural and Multidisciplinary
  Optimization (WCSMO3)}, 1999.

\bibitem{Fritzen2016Topology}
Felix Fritzen, Liang Xia, Matthias Leuschner, and Piotr Breitkopf.
\newblock Topology optimization of multiscale elastoviscoplastic structures.
\newblock {\em International Journal for Numerical Methods in Engineering},
  106:430--453, 2016.

\bibitem{harzheim2008strukturoptimierung}
Lothar Harzheim.
\newblock Strukturoptimierung.
\newblock {\em Harri Deutsch, Frankfurt}, 2008.

\bibitem{Hencky1924ZurTheorie}
Heinrich Hencky.
\newblock Zur theorie plastischer deformationen und der hierdurch im material
  hervorgerufenen nachspannungen.
\newblock {\em ZAMM - Zeitschrift für Angewandte Mathematik und Mechanik},
  4(4):323--334, 1924.

\bibitem{Hencky1933TheNewTheory}
Heinrich Hencky.
\newblock The new theory of plasticity, strain hardening, and creep, and the
  testing of the inelastic behavior of metals.
\newblock {\em Journal of Applied Mechanics}, 1(4):151--155, 1933.

\bibitem{Huang2008Topology}
X.~Huang and Y.~M. Xie.
\newblock Topology optimization of nonlinear structures under displacement
  loading.
\newblock {\em Engineering Structures}, 30(7):2057--2068, 2008.

\bibitem{jantos2019accurate}
Dustin~R. Jantos, Klaus Hackl, and Philipp Junker.
\newblock An accurate and fast regularization approach to thermodynamic
  topology optimization.
\newblock {\em International Journal for Numerical Methods in Engineering},
  117(9):991--1017, 2019.

\bibitem{junker2021extended}
Philipp Junker and Daniel Balzani.
\newblock An extended hamilton principle as unifying theory for coupled
  problems and dissipative microstructure evolution.
\newblock {\em Continuum Mechanics and Thermodynamics}, 33(4):1931--1956, 2021.

\bibitem{Junker2021hyperelastic}
Philipp Junker and Daniel Balzani.
\newblock A new variational approach for the thermodynamic topology
  optimization of hyperelastic structures.
\newblock {\em Computational Mechanics}, (67):455--480, 2021.

\bibitem{junker2015variational}
Philipp Junker and Klaus Hackl.
\newblock A variational growth approach to topology optimization.
\newblock {\em Structural and Multidisciplinary Optimization}, 52(2):293--304,
  2015.

\bibitem{junker2017numerical}
Philipp Junker and Philipp Hempel.
\newblock Numerical study of the plasticity-induced stabilization effect on
  martensitic transformations in shape memory alloys.
\newblock {\em Shape Memory and Superelasticity}, 3(4):422--430, 2017.

\bibitem{Kick2022Implementation}
Miriam Kick, Dustin~R. Jantos, and Philipp Junker.
\newblock {Dataset: Implementation of thermodynamic topology optimization for
  hardening materials in Julia}, 2022.
\newblock https://doi.org/10.25835/ya8glznn.

\bibitem{Li2017Design}
Lei Li, Guodong Zhang, and Kapil Khandelwal.
\newblock Design of energy dissipating elastoplastic structures under cyclic
  loads using topology optimization.
\newblock {\em Structural and Multidisciplinary Optimization}, 56(2):391--412,
  2017.

\bibitem{Li2017Topology}
Lei Li, Guodong Zhang, and Kapil Khandelwal.
\newblock Topology optimization of energy absorbing structures with maximum
  damage constraint.
\newblock {\em International Journal for Numerical Methods in Engineering},
  112:737--775, 2017.

\bibitem{Luo2012Topology}
Yangjun Luo and Zhan Kang.
\newblock Topology optimization of continuum structures with drucker–prager
  yield stress constraints.
\newblock {\em Computers and Structures}, 90--91:65--75, 2012.

\bibitem{Maury2018ElastPlastic}
Aymeric Maury, Grégoire Allaire, and François Jouve.
\newblock Elasto-plastic shape optimization using the level set method.
\newblock {\em SIAM Journal on Control and Optimization}, 56(1):556--581, 2018.

\bibitem{Maute1998Adaptive}
K.~Maute, S.~Schwarz, and E.~Ramm.
\newblock Adaptive topology optimization of elastoplastic structures.
\newblock {\em Structural Optimization}, 15:81--91, 1998.

\bibitem{Nakshatrala2015Topology}
P.~B. Nakshatrala and D.~A. Tortorelli.
\newblock Topology optimization for effective energy propagation in
  rate-independent elastoplastic material systems.
\newblock {\em Computer Methods in Applied Mechanics and Engineering},
  295:305--326, 2015.

\bibitem{RambergOsgood1943Description}
Walter Ramberg and William~R. Osgood.
\newblock Description of stress-strain curves by three parameters.
\newblock NTRS - NASA Technical Reports Server(NACA-TN-902), 1943.

\bibitem{Russ2021ANovel}
Jonathan~B. Russ and Haim Waisman.
\newblock A novel elastoplastic topology optimization formulation for enhanced
  failure resistance via local ductile failure constraints and linear buckling
  analysis.
\newblock {\em Computer Methods in Applied Mechanics and Engineering},
  373:113478, 2021.

\bibitem{schumacher2013Optimierung}
Axel Schumacher.
\newblock {\em Optimierung mechanischer Strukturen: Grundlagen und industrielle
  Anwendungem}.
\newblock Springer, 2013.

\bibitem{Schwarz2001Topology}
Stefan Schwarz, Kurt Maute, and Ekkehard Ramm.
\newblock Topology and shape optimization for elastoplastic structural
  response.
\newblock {\em Computer Methods in Applied Mechanics and Engineering},
  190(15-17):2135--2155, 2001.

\bibitem{sigmund2013topology}
Ole Sigmund and Kurt Maute.
\newblock Topology optimization approaches: {A} comparative review.
\newblock {\em Structural and Multidisciplinary Optimization},
  48(6):1031--1055, 2013.

\bibitem{Swan1998VoigtReuss}
C.~Swan and I.~Kosaka.
\newblock Voigt–reuss topology optimization for structures with nonlinear
  material behaviors.
\newblock {\em International Journal for Numerical Methods in Engineering},
  40:3785--3814, 1998.

\bibitem{vogel2020adaptive}
Andreas Vogel and Philipp Junker.
\newblock Adaptive thermodynamic topology optimization.
\newblock {\em Structural and multidisciplinary optimization}, accepted for
  publication, 2020.

\bibitem{Wallin2016Topology}
Mathias Wallin, Viktor Jönsson, and Eric Wingren.
\newblock Topology optimization based on finite strain plasticity.
\newblock {\em Struct Multidisc Optim}, 54:783--793, 2016.

\bibitem{wriggers2008nonlinear}
Peter Wriggers.
\newblock {\em Nonlinear finite element methods}.
\newblock Springer Science \& Business Media, 2008.

\bibitem{Xia2017Evolutionary}
Liang Xia, Felix Fritzen, and Piotr Breitkopf.
\newblock Evolutionary topology optimization of elastoplastic structures.
\newblock {\em Structural and Multidisciplinary Optimization}, 55:569--581,
  2017.

\bibitem{Yoon2007Topology}
Gil~Ho Yoon and Yoon~Young Kim.
\newblock Topology optimization of material-nonlinear continuum structures by
  the element connectivity parameterization.
\newblock {\em International Journal for Numerical Methods in Engineering},
  69(10):2196--2218, 2007.

\bibitem{Yuge1995Optimization}
K.~Yuge and N.~Kikuchi.
\newblock Optimization of a frame structure subjected to a plastic deformation.
\newblock {\em Structural Optimization}, 10:197--2018, 1995.

\bibitem{Zhang2017Topology}
Guodong Zhang, Lei Li, and Kapil Khandelwal.
\newblock Topology optimization of structures with anisotropic plastic
  materials using enhanced assumed strain elements.
\newblock {\em Structural and Multidisciplinary Optimization},
  55(6):1965--1988, 2017.

\bibitem{Zhao2020Topology}
Tuo Zhao, Eduardo~N. Lages, Adeildo~S. Ramos~Jr., and Glaucio~H. Paulino.
\newblock Topology optimization considering the drucker–prager criterion with
  a surrogate nonlinear elastic constitutive model.
\newblock {\em Structural and Multidisciplinary Optimization}, 62:3205--3227,
  2020.

\bibitem{Zhao2019Material}
Tuo Zhao, Adeildo~S. Ramos~Jr., and Glaucio~H. Paulino.
\newblock Material nonlinear topology optimization considering the von mises
  criterion through an asymptotic approach: Max strain energy and max load
  factor formulations.
\newblock {\em International Journal for Numerical Methods in Engineering},
  118:804--828, 2019.

\end{thebibliography}
